\documentclass[twocolumn,epsf,showpacs,amsmath,amssymb,floatfix,pre]{revtex4}
\usepackage{bm}
\usepackage{graphicx}
\usepackage{amsmath}
\usepackage{dcolumn}% Align table columns on decimal point

\begin{document}

\title{Effect of pressure on the phase behavior and structure of
water confined between nanoscale hydrophobic and hydrophilic
plates}

\author{Nicolas Giovambattista$^1$, Peter J. Rossky$^2$, and Pablo G. Debenedetti$^1$}

\address{
$^1$ Department of Chemical Engineering,\\Princeton University, Princeton, NJ 08544-5263 USA\\
$^2$ Department of Chemistry and Biochemistry, Institute for Theoretical Chemistry,\\
University of Texas at Austin, Austin, Texas 78712 USA
}

\date{17 Feb. 2006}

%\affiliation{Department of Chemical Engineering\\Princeton University, Princeton, NJ 08544}

\begin{abstract}
We perform systematic molecular dynamics simulations of water
confined between two nanoscale plates at $T=300$~K. We investigate
the effect of pressure ($-0.15$~GPa~$\leq P \leq 0.2$~GPa) and
plate separation ($0.4$~nm~$\leq d \leq 1.6$~nm) on the phase
behavior of water when the plates are either hydrophobic or
hydrophilic. When water is confined between hydrophobic plates,
capillary evaporation occurs between the plates at low enough $P$.
The threshold value of $d$ at which this transition occurs
decreases with $P$ (e.g., $1.6$~nm at $P \approx -0.05$~GPa;
$0.5$~nm at $P \approx 0.1$~GPa), until, at high $P$, no capillary
evaporation occurs. For $d \approx 0.6$~nm and $P \geq 0.1$~GPa,
the system crystallizes into a bilayer ice.
 A $P-d$ phase diagram showing the vapor, liquid, and bilayer ice phases
 is proposed. When water
is confined by hydrophilic (hydroxylated silica) plates, it
remains in the liquid phase at all $P$ and $d$ studied.
Interestingly, we observe, for this case, that even at the $P$ at
which bulk water cavitates, the confined water remains in the
liquid state. We also study systematically the state of hydration
at different $P$ for both kinds of plates. For the range of
conditions studied here, we find that in the presence of
hydrophobic plates the effect of $P$ is to enhance water structure
and to push water molecules toward the plates. The average
orientation of water molecules next to the hydrophobic plates does
not change upon pressurization. In contrast, in the presence of
hydrophilic plates, water structure is insensitive to $P$. Hence,
our results suggest that upon pressurization, hydrophobic plates
behave as `soft' surfaces (in the sense of accommodating
pressure-dependent changes in water structure) while hydrophilic
walls behave as `hard' surfaces.
\end{abstract}

\maketitle

\section{Introduction}

Confining geometries that contain thin water films are common in
biology, geology and engineering. Examples include ion channels
\cite{T6,beckstein,taj,murata,fu,sui}, mesoscopic surfactant
assemblies \cite{T5,ladanyi}, mineral inclusions \cite{pabloBook},
zeolites \cite{T1}, and microfluidic technologies \cite{troian}.
Understanding the changes in water structure, dynamics and
thermodynamics due to interactions with confining surfaces is
therefore of interest in a wide variety of scientific problems and
technical applications, such as the inhibition of corrosion
\cite{bookCorrosion}, heterogeneous catalysis, the design of
superhydrophobic surfaces \cite{superHphob}, the ascent of sap in
plants \cite{pabloBook}, the function of biological membranes
\cite{pratt02}, and the development of `lab on a chip'
technologies \cite{T12}.

The behavior of water near hydrophobic surfaces has attracted
considerable attention, both on account of the intrinsic
scientific interest of the thermodynamic problem (e.g.,
\cite{tanford,ben,israel,chandler2,lcw,still73,hummer3,truskett01,southall,widom,robinson}),
as well as its relevance to water permeation through membrane
channels \cite{beckstein,taj,murata,fu,sui,hinds} and carbon
nanotubes \cite{hummerTube,andreev}, micelle formation
\cite{hunter,israel}, and the amphiphaticity of membrane proteins
\cite{cheng,hummer0,pratt02}. Hydrophobic interactions are also
important in phase transfer catalysis \cite{dogo}, chemical
self-assembly of macroscopic objects \cite{terfort}, and capillary
evaporation \cite{luzar,luzar2,rowlinson}.

More than thirty years ago, Stillinger \cite{still73} argued that
when a non-polar solute is sufficiently large, the water-solute
interface resembles that between water and its own vapor. Lum {\it
et al.} \cite{lcw} proposed a theoretical approach to describe the
crossover between the solvation of small and large hydrophobic
objects. Bulk thermodynamics and entropic contributions to the
free energy of solvation are dominant in the former case.
Interfacial thermodynamics and enthalpic contributions to
solvation free energy play a key role in the latter case. The
crossover length scale between these two regimes, $\approx 1$~nm
at ambient conditions, is a sensitive function of pressure and the
presence of co-solutes \cite{truskett}. From a microscopic point
of view, this length scale is related to the different manner in
which water molecules arrange around solutes of different sizes.
In the presence of small hydrophobic solutes, water molecules
rearrange in such a way that
 the solutes can be accommodated into water's hydrogen-bond (HB)
network. Thus, small hydrophobic solutes are caged in
clathrate-like structures \cite{zichi,du,margulis1} with HB
vectors (defined as the four tetrahedral vectors pointing outward
from the oxygen atom of each water molecule along the two
oxygen-hydrogen bond donor, and the two ``lone pair electrons''
acceptor directions) avoiding pointing toward the non-polar
solute. As the solute size increases (to $\approx 1$~nm at ambient
conditions) \cite{belch,wall91,wall95,wall95b}, such
clathrate-like structure cannot be maintained: HBs break, inducing
a restructuring of the HB network. In this case, the hydration
structure shows an orientational inversion with respect to the
clathrate-like structure, with HBs pointing toward the solute
\cite{lee1,belch,torrie,wall91,gardner}.

When two hydrophobic surfaces approach each other, a confined
water film can be thermodynamically destabilized with respect to
the vapor at small enough separations
\cite{truskett01,lcw,luzar,luzar2,pettittJACS}. A plausible
interpretation of the experimentally observed long-ranged
attractions between hydrophobic surfaces (e.g., $10-100$~nm)
\cite{parker,tsao,pashley,claeson} involves dewetting of the
inter-plate region. As shown by B\'{e}rard et al. \cite{patey88},
such dewetting is not particular to water, but arises as a general
consequence of weak fluid-wall interactions relative to
fluid-fluid interactions \cite{truskett01}. The generality of
confinement-induced drying notwithstanding, interest in this
phenomenon has focused predominantly on the case in which the
confined fluid is water and the confining surfaces are
hydrophobic. Several computational studies of this situation exist
(e.g.,
\cite{luzar2,jensen1,wall95,margulis2,huang1,graphite,koishi,gordillo,pettittJCPB}).

Kauzmann's influential review \cite{kauz} first pointed out the
importance of hydrophobic interactions in protein folding. It is
generally accepted that the formation of a core of hydrophobic
residues from which water is partially excluded is a dominant
force in the folding of globular proteins
\cite{kauz,dill,fersht,brooks,dobson,privalov,honig,sturt,williams,froloff}.
However, the surface of a protein is a geometrically, chemically,
and electrically heterogeneous object, and the distribution,
structure, and dynamics of water near such a surface is expected
to be quite different from that found near idealized surfaces,
such as perfectly hydrophobic walls. In fact, computer simulations
have clearly shown a range of water behavior in response to
surface heterogeneity. Cheng and Rossky \cite{rosskyNat} studied
the hydration of the polypeptide melittin. They found that
clathrate-like structures dominate near convex surface patches,
while near flat regions the hydration shell fluctuates between
clathrate-like and less ordered (inverted) structures. The
collapse of a two-domain BphC enzyme showed no dewetting in the
inter-domain region when the units were brought together down to a
distance of $0.4$~nm \cite{margulis3}. A dewetting transition was
observed only when electrostatic interactions were turned off
\cite{margulis3}. In contrast, dewetting has been observed in the
collapse of of the melittin tetramer \cite{berneNew}; however,
even single mutations were shown to prevent drying.

The above protein folding examples illustrate the importance of
understanding water behavior near complex, heterogeneous surfaces.
Materials science and engineering applications involving
chemically nanopatterned substrates (e.g., \cite{dePablo})
likewise will require an understanding of water structure near
heterogeneous interfaces. Molecular simulations should prove
extremely powerful in addressing this problem, because they allow
exquisitely sensitive control of surface geometry and chemistry.
Questions such as the manner in which the characteristic length of
hydrophobic patches and the pattern of their distribution on a
hydrophilic surface affect confined water structure and dynamics,
or the effect of surface chemistry in promoting or preventing bulk
cavitation following dewetting of a confined region are ideally
suited for computational scrutiny. The recent work of Koishi {\it
et al.} \cite{koishi} is an excellent example of the valuable
insights that can be obtained by molecular simulations regarding
water behavior near complex surfaces.

This paper is the first report of an ongoing computational
investigation aimed at understanding the influence of surface
heterogeneity and bulk thermodynamic conditions (pressure,
temperature) on the structure, dynamics, and thermodynamics of
confined water. Here we establish the `base case' conditions by
comparing water structure and phase behavior at $300$~K when
confined by purely hydrophobic and purely hydrophilic nanoscale
surfaces ($0.4 \leq d \leq 1.6$~nm). Li {\it et al.}
\cite{graphite} have recently studied hydration and dewetting near
hydrophilic and hydrophobic plates (but with different chemistry
than the ones we consider here), at a single pressure. An
important aspect of the present work is that we perform such a
comparison over a broad range of pressures ($-0.15 \leq P \leq
0.2$~GPa). In subsequent papers we will report results on water
confined between heterogeneous plates with engineered
`patchiness'.

Our computational study should be placed in the context of a
considerable body of experimental work aimed at studying
systematically the properties of water confined by hydrophobic and
hydrophilic surfaces. The forces between hydrophobic surfaces in
aqueous solution have been measured with the surface force
apparatus (e.g., \cite{uno,pashley,claeson,tsao,parker}). Such
studies revealed the existence of long-range attractions between
hydrophobic surfaces, over distances of the order of 100 nm, and
investigated the origin of this phenomenon. Bridging cavities and
microbubbles were linked in several \cite{pashley,claeson,parker},
but not all \cite{tsao} of these studies, to the long-ranged
attraction. Zhang {\it et al.} \cite{seis} investigated the
response to shear stresses of water confined between adjoining
surfaces, one hydrophobic and the other hydrophilic (Janus
surface). They found that the competing effects of these surfaces
gave rise to unusually noisy responses to shear. Jensen {\it et
al.} \cite{jensen1} used X-ray reflectivity to study the contact
region between water and an extended paraffin surface. They found
that drying was confined to a very narrow ($\approx 1.5$~nm)
region. Ruan {\it et al.} \cite{ocho} investigated the structure
and dynamics of water near a hydrophilic surface using ultrafast
electron crystallography. They observed coexistence between
ordered surface water (up to $1$~nm thick) and crystalline
islands. Reference to particular aspects of the studies summarized
above will be made throughout this paper, stressing, whenever
possible, the relationship between experimental observations and
our calculations.

This paper is organized as follows. In the next section we
describe the simulation details. The results for hydrophobic
plates are presented in Sec.~\ref{Hphobic}; calculations on water
confined by hydrophilic plates are reported in Sec.~\ref{Hphilic}.
We summarize our main conclusions in Sec.~\ref{summary}.

\section{Simulation details}
\label{simulations}

We perform molecular dynamics (MD) simulations in the NPT
ensemble. The system is composed of $N=3375$ water molecules in
which are inmersed two identical finite three dimensional
nanoscale plates. The temperature is fixed at $T=300$ K by using a
Berendsen thermostat \cite{berendsenThermo}. The pressure ranges
from $P=-0.15$~GPa to $P=0.2$~GPa, and is controlled by coupling
the system volume to an external bath at $P$ (analogous to the
Berendsen thermostat \cite{berendsenThermo}).

\begin{figure}
\centering{
    \includegraphics[width=2.5 in]{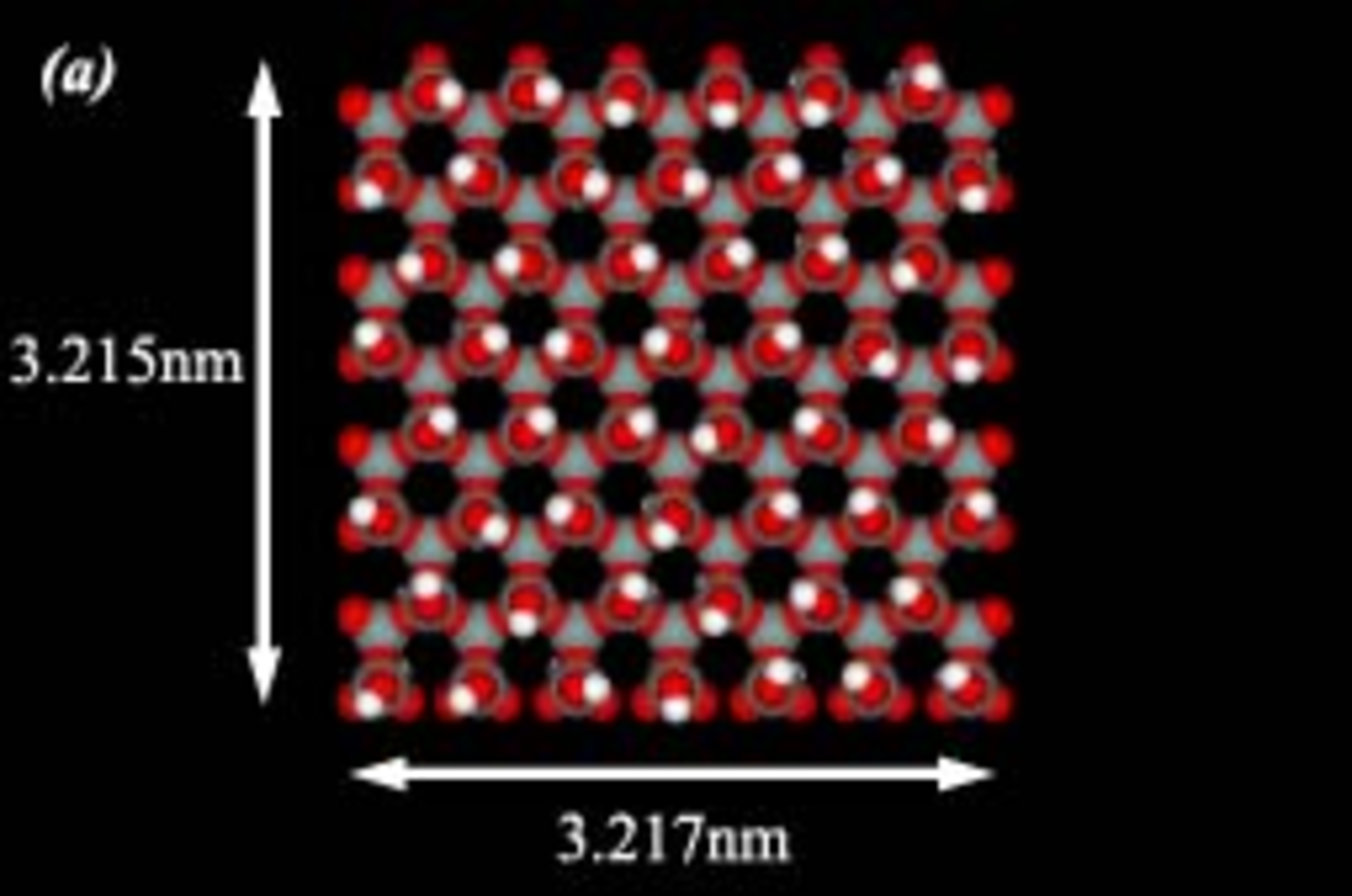}
    \includegraphics[width=2.5 in]{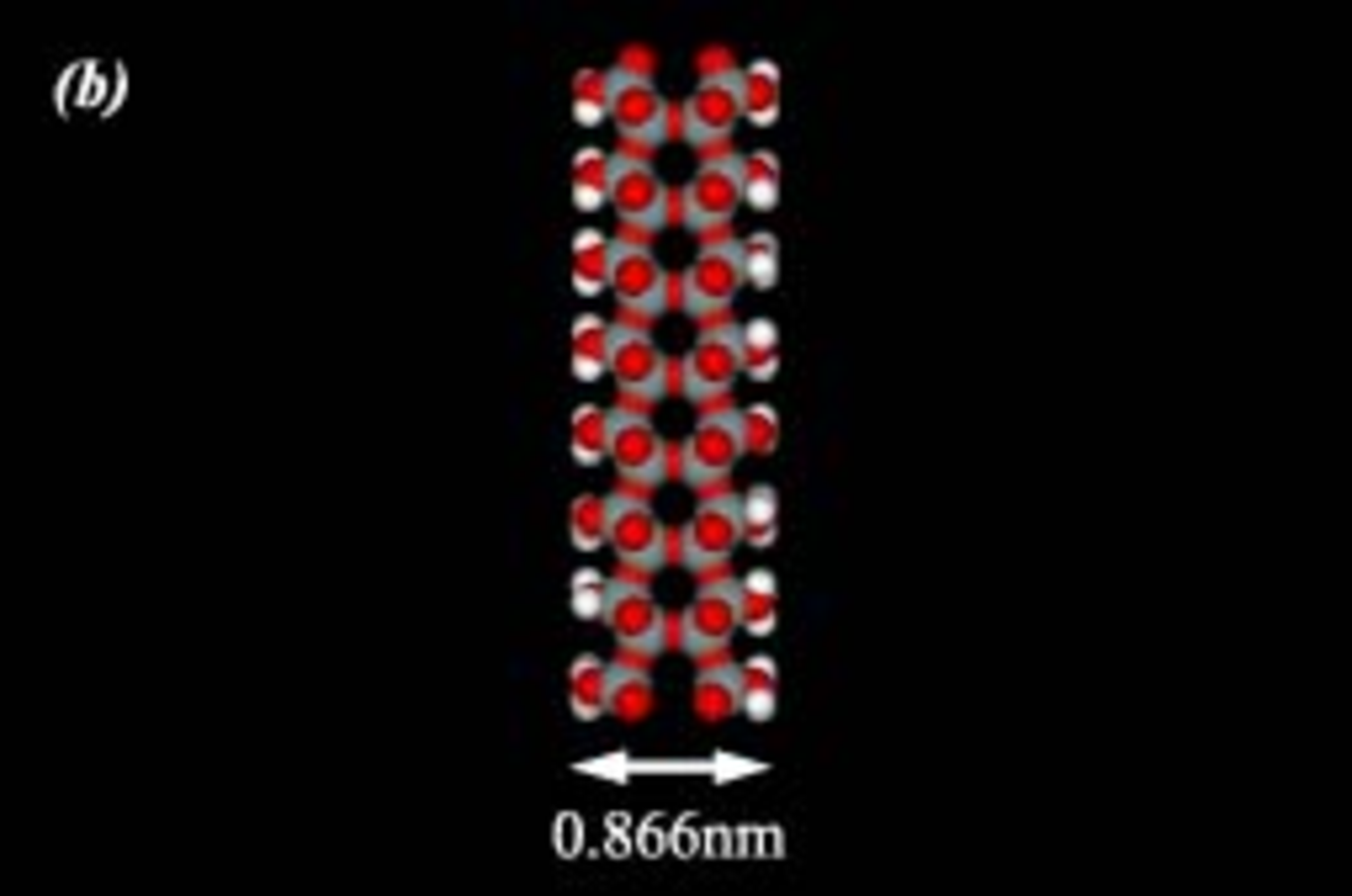}
    }
\centering{
    \includegraphics[width=2.5 in]{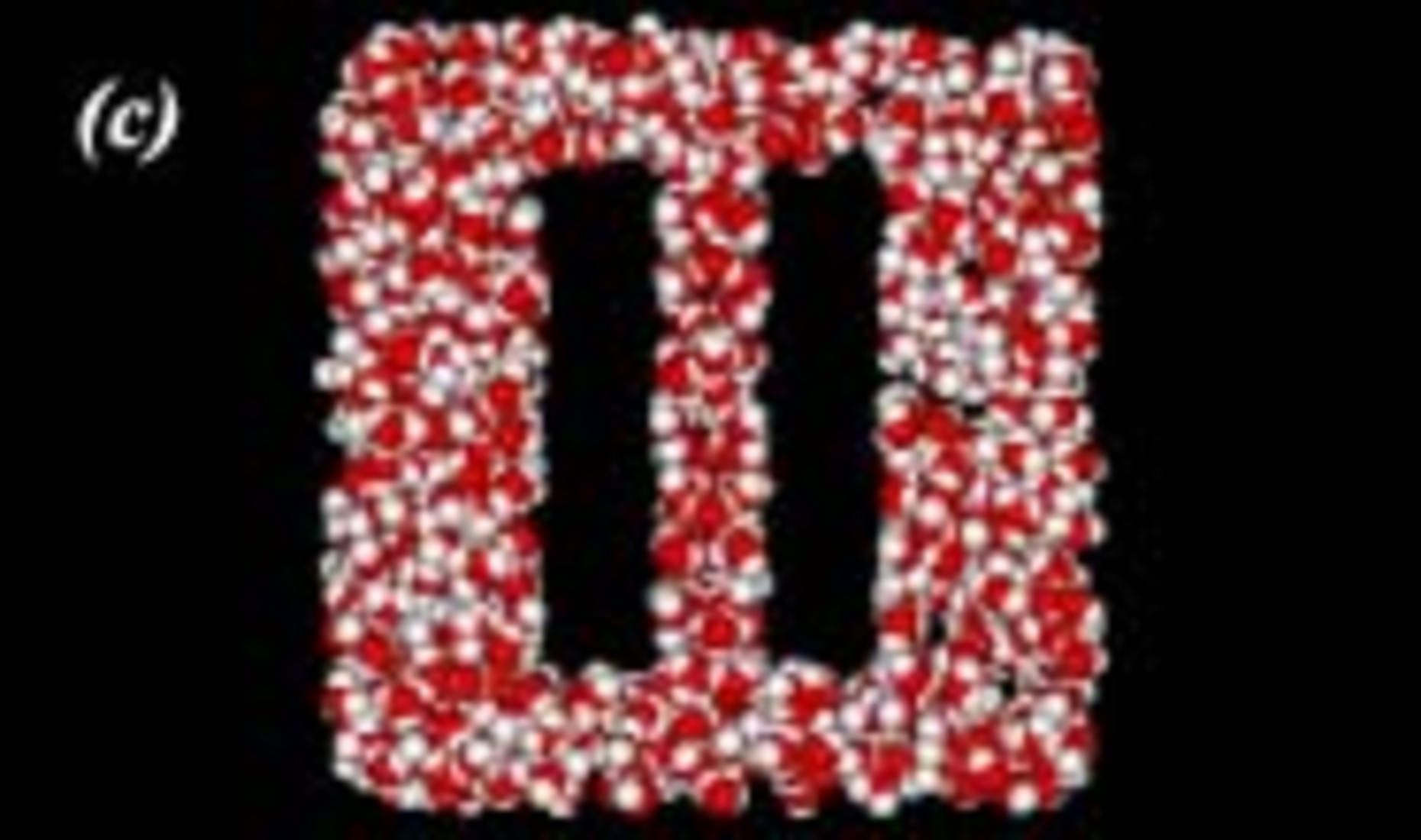}
    }
\centering{
    \includegraphics[width=2.5 in]{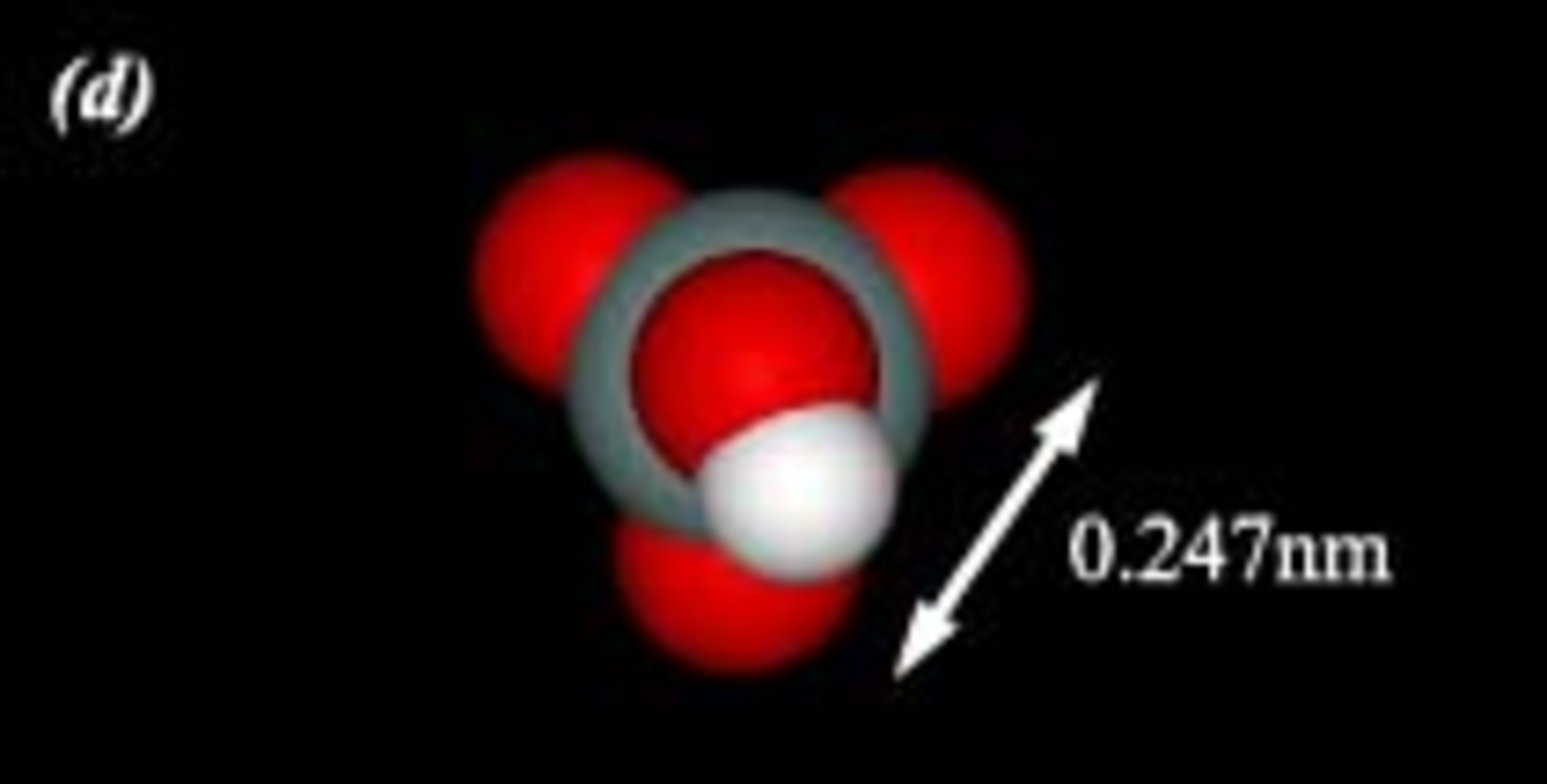}
    \includegraphics[width=2.5 in]{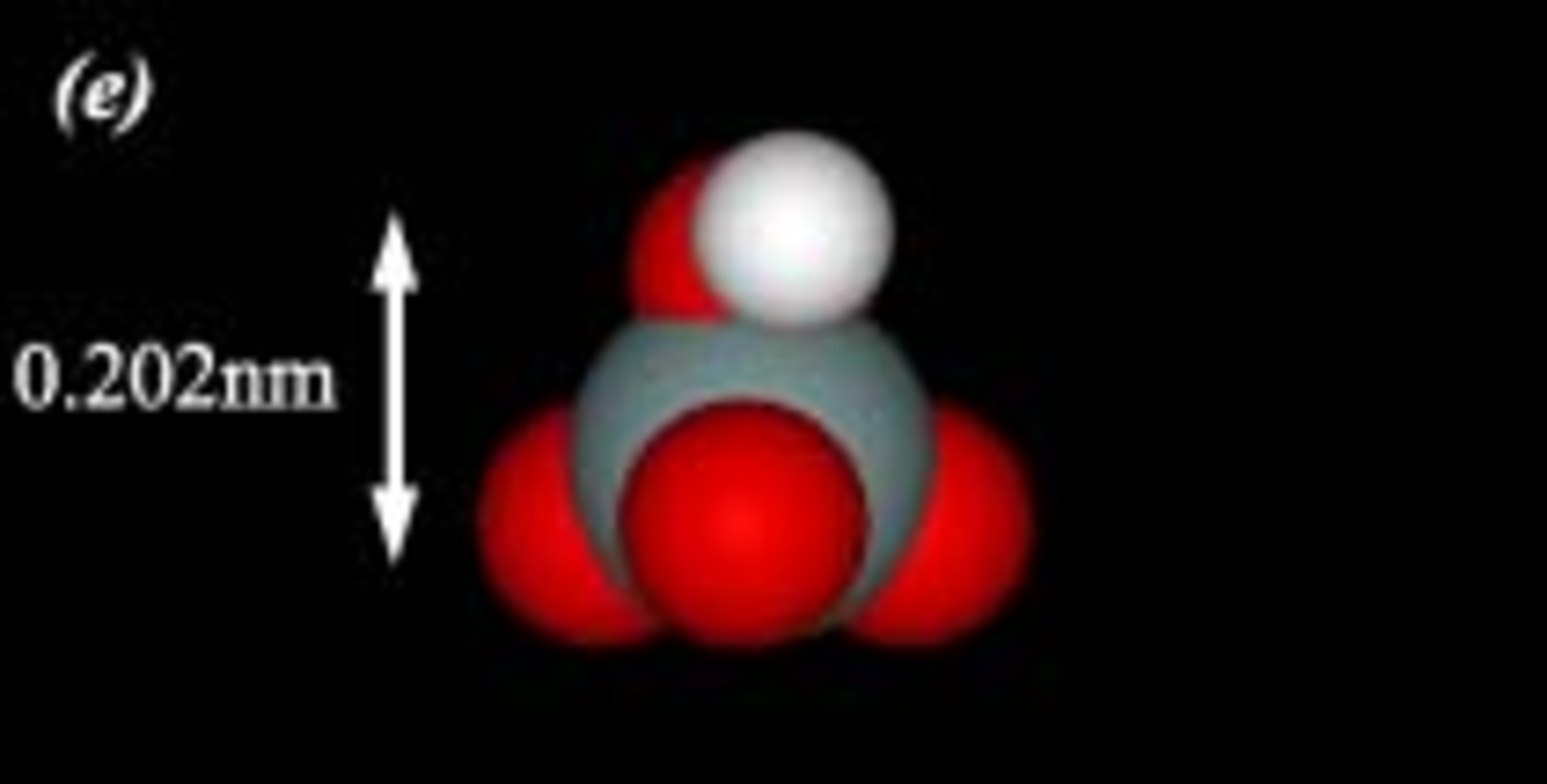}
    }
\caption{(Color online) (a) Front and (b) lateral view of the
fully hydroxylated silica plate [(1.1.1) octahedral face of
cristobalite], one of the two kinds of plates used in this study.
White, red, and gray spheres represent hydrogen, oxygen, and
silicon atoms, respectively. (c) $y-z$ cross section showing the
space available to the water molecules. (d) Top and (e) lateral
view of the surface $SiO_4$ unit, in which the surface oxygen atom
is hydroxylated. } \label{SnapshotWalls}
\end{figure}

We simulate a cubic sample using periodic boundary conditions
along the three directions. Water molecules are modelled with the
SPC/E pair potential \cite{berensen}. The plates are introduced
symmetrically about the center of the box such that they are
parallel to the $x-y$ plane and equidistant from the $z=0$ plane.
The plate dimensions ($3.215 \times 3.217 \times 0.866$ nm$^3$)
are smaller than the box size (which varies with $P$ but always
exceeds $4.85$~nm under the conditions investigated here).
 Figures \ref{SnapshotWalls}(a) and (b) show a fully hydroxylated silica plate
  (one of the two kinds of plates used in this study) and
Fig. \ref{SnapshotWalls}(c) is a cross section, showing the water
molecules and the two cavities where the plates are located. The
plates are fixed in space throughout.

We consider two kinds of plates: hydrophobic or hydrophilic. Their
common underlying structure corresponds to four layers of $SiO_2$
 [see Fig. \ref{SnapshotWalls}(b)] reproducing the
 $(1.1.1)$ octahedral face of cristobalite \cite{books,website}
 [see Fig. \ref{SnapshotWalls}(a)].
 The unit cell of $SiO_2$ is idealized as a perfect tetrahedron with
 $O-O$ and $Si-O$ distances of $0.247$~nm and $0.151$~nm, respectively.
The hydrophobic plates consist of the above-described structure;
however, the `$Si$' and `$O$' atoms interact with water molecule
$O$ atoms exclusively via a Lennard-Jones potential ($\epsilon$
and $\sigma$ parameters are given in Table \ref{tabla}).
 Each hydrophobic plate is composed of $674$ atoms.

The hydrophilic plates correspond to fully hydroxylated silica and
 are obtained by
attaching a hydrogen atom to each surface oxygen atom on the four
plate surfaces [see Figs. \ref{SnapshotWalls}(d) and (e)]. The
$O-H$ distance is chosen to be the same as in the SPC/E model,
i.e. $0.1$ nm. The $Si$ and $O$ atoms are located in fixed
positions (as in the hydrophobic plates)
 but the $H$ atoms on the surface are able to move with fixed
  bond lengths and bond angles; each $H$ atom can
reorient in a circle. Such circular motion occurs in a plane
parallel to the plate at a distance $0.033$ nm away from the $O$
atom plane of the $Si-O-H$ groups. The resulting $Si-O-H$ angle is
$109^o.27$ [see Fig. \ref{SnapshotWalls}(e)]. Each hydrophilic
plate is composed of $778$ atoms. Only the polar $Si-O-H$ groups
at the surface carry partial electric charge [charge values are
given in Table \ref{tabla}]. Electrostatic interactions are
treated using the Ewald sum method with a cutoff distance of
$0.79$ nm and parameters $m_{max}=5^3$ (for the number of vectors
in the reciprocal-space sum) and $\alpha=0.4$ (for the width of
the screening-charge Gaussian distribution) \cite{ewald,virial}.

The vibrations of the $Si$ and $O$ atoms have not been taken into
account in this work. A computational study of water in
single-wall carbon nanotubes indicates that flexibility can affect
a channel's apparent hydrophobicity \cite{andreev}. On the other
hand, a molecular dynamics study of water droplets on graphite
revealed a negligible effect of substrate vibrations on the
contact angle \cite{werder}. The lattice constant for the
cristobalite structure used in this work is $0.494$~nm. Invoking a
Lindemann-type estimate, the amplitude of individual atomic
vibrations should be considerably smaller than the melting
threshold ($10 \%$ of the lattice constant), i.e. $0.05$~nm in our
case. This upper bound on the magnitude of substrate atomic
vibrations is an order of magnitude smaller than the closest
inter-plate separation considered in our work. Thus, while the
effect of crystal vibrations on the structure and dynamics of
confined water deserves careful attention and  will be explored in
future studies, we believe that the  present rigid wall base case
is a reasonable starting point for our investigations.

The pressure is calculated using the virial expression, taking
into account the fact that some atoms in the plates are fixed and
others can move \cite{virial,footnoteP}. We also use the link cell
and neighbor list methods to calculate the pair interactions
\cite{rappaport}.

We perform simulations for different time intervals depending on
the kind of plates and their separation. Simulation times are
indicated in Table~\ref{tabla2}. One of the reasons we chose these
simulation times is that when there is no capillary evaporation or
crystallization, we find that quantities such us total energy and
volume are constant for $t > 50$~ps. Thus, the first $50$~ps of
the simulation are discarded and the rest of the simulation is
used for data acquisition. Moreover, the correlation time
(obtained from the intermediate scattering function) in bulk water
simulations using the SPC/E model at $T=300$~K and
$\rho=1.0$~g/cm$^3$ is approximately $3$~ps \cite{francislong}.
Thus, our simulations appear long enough to avoid effects of the
equilibration process on our calculations.

\section{Results: Hydrophobic plates}
\label{Hphobic}

\subsection{Capillary evaporation: effect of pressure}
\label{P-d-Hphobic}

In this section, we describe the effects of pressure $P$ and
plate-plate separation $d$ on the phase behavior of water confined
between hydrophobic plates at $T=300$~K.
 To consistently define $d$, we use the
distance between the planes containing the hydrogen ($H$) atoms on
the inner surfaces of two hydrophilic plates. In the case of
hydrophobic plates, where no hydrogen atoms are present, the same
planes (where the $H$ atoms would hypothetically be located) are
used to define $d$. As previously noted, the plane containing the
$H$ atoms is located at a distance $0.033$ nm from the plane
containing the oxygen atoms of the silanol groups.

\begin{figure}
\includegraphics[width=4.0 in]{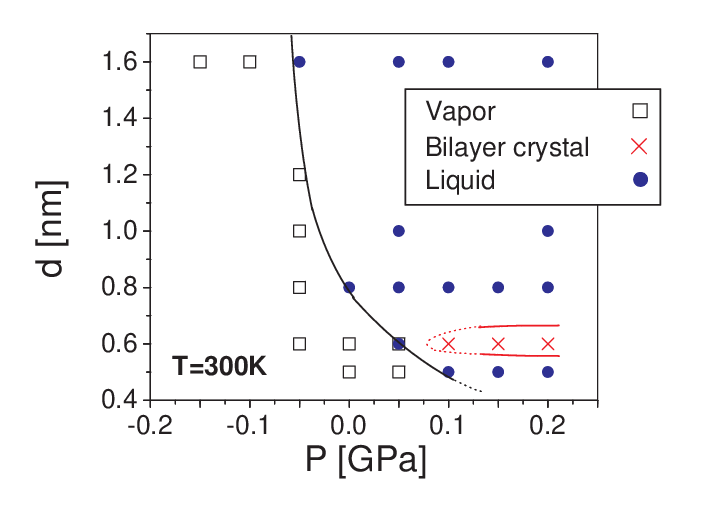}
\caption{(Color online) Phase behavior of confined water as a
function of pressure and separation between nanoscale hydrophobic
plates. The temperature is fixed at $T=300$~K. Filled circles
indicate liquid state points.
 Empty squares indicate vapor state points at
which capillary evaporation occurs (i.e., we observe no water
molecule between plates). For $d \approx 0.6$~nm, a bilayer
crystal is formed as indicated by red $\times$ symbols. At
$d=0.6$~nm and $P=0.05$~GPa, fluctuations between liquid and vapor
states are observed for at least $1$~ns. Continuous lines suggest
a schematic phase diagram based on the simulation results. In the
figure, the dashed red line corresponds to the assumption that the
crystal region is not connected with the liquid-vapor transition
line, i.e. that there is no `triple point' between the liquid, the
vapor, and the bilayer crystal. The black dotted line is the
extrapolation of the liquid-vapor transition line into the $d \leq
0.4$~nm region, where no liquid can be simulated due to the small
distance between plates.} \label{PhaseDiag-d-P}
\end{figure}

Figure \ref{PhaseDiag-d-P} summarizes the results of our MD
simulations for $d \geq 0.5$~nm.
 Two observations are most relevant from Fig.~\ref{PhaseDiag-d-P}: (i) the
 threshold separation distance $d_{th}$ required for capillary evaporation
 decreases with increasing $P$, and (ii) for $d \approx 0.6$~nm a bilayer crystal is
 obtained at high $P$.
Simulations for $d \leq 0.4$~nm at $P=0.2$~GPa and $0$~GPa show
that molecules are trapped interstitially between the plate atoms.
The water molecules remain in a plane and are not able to diffuse
due to the lack of space between plates.

The decrease in $d_{th}$ with increasing $P$ can be understood as
follows. Equating the grand potential of the confined liquid and
the confined vapor, for sufficiently large surfaces
\cite{luzar,parker},
\begin{equation}
- ~ P ~ A ~ d_{th} ~+~ 2 ~ A ~ \gamma_{wl}  \approx
 - ~ P^* ~ A ~d_{th}~+~ 2 ~ A ~ \gamma_{wv}
\end{equation}
whence \cite{luzar,parker,patey88},
\begin{equation}
d_{th} \approx \frac{2 ~ \Delta \gamma}{ (P-P^*)} \label{dth}
\end{equation}
In the above equations, $A$ is the plate surface area;
$\gamma_{wl}$ and $\gamma_{wv}$ are the wall-liquid and wall-vapor
interfacial tensions ($\Delta \gamma \equiv \gamma_{wl} -
\gamma_{wv}$); $P$ is the bulk pressure; and $P^*$ is the
equilibrium vapor pressure at the given temperature. It follows
from Eq.~(\ref{dth}) that $d_{th}$ decreases with increasing $P$,
as manifest in Fig.~\ref{PhaseDiag-d-P}. For an incompressible
liquid, $P-P^*= \rho_l~ (\mu_l-\mu_v)$, and, therefore
\cite{lcw,luzar},
\begin{equation}
d_{th} \approx  \frac{2 ~ \Delta \gamma}{ \rho_l ~ (\mu_l -
\mu_v)} \label{dth2}
\end{equation}
where $\mu_l$ and $\mu_v$ denote the chemical potentials of the
liquid at $P$ and of the vapor (and liquid) at $P^*$. In the above
derivation we have assumed that the plate characteristic
size,~$\approx A^{1/2}$, is large enough so that one can neglect
the finite lateral size of the confined region (i.e.,
$d_{th}/A^{1/2} \ll 1$) \cite{luzar}.

\begin{figure}
\centering{
    \includegraphics[width=3 in]{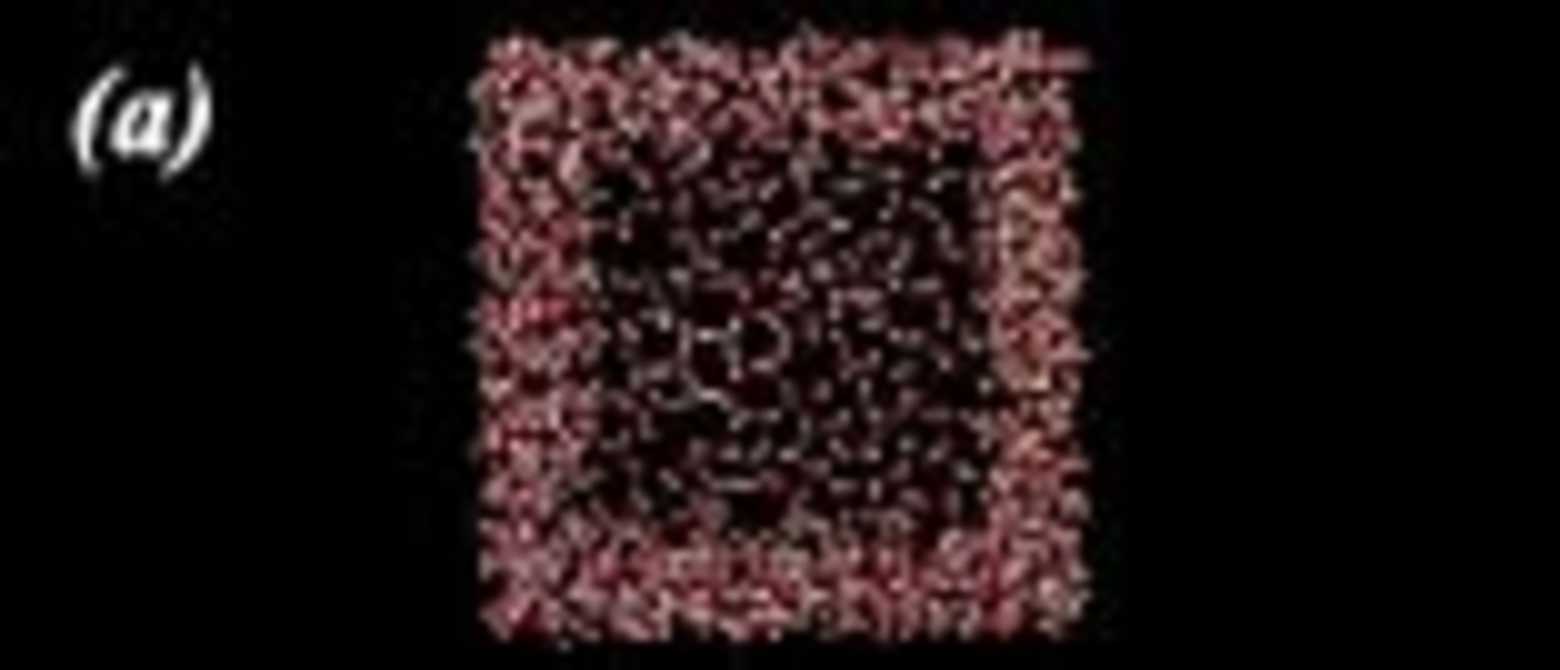}
    \includegraphics[width=3.0in]{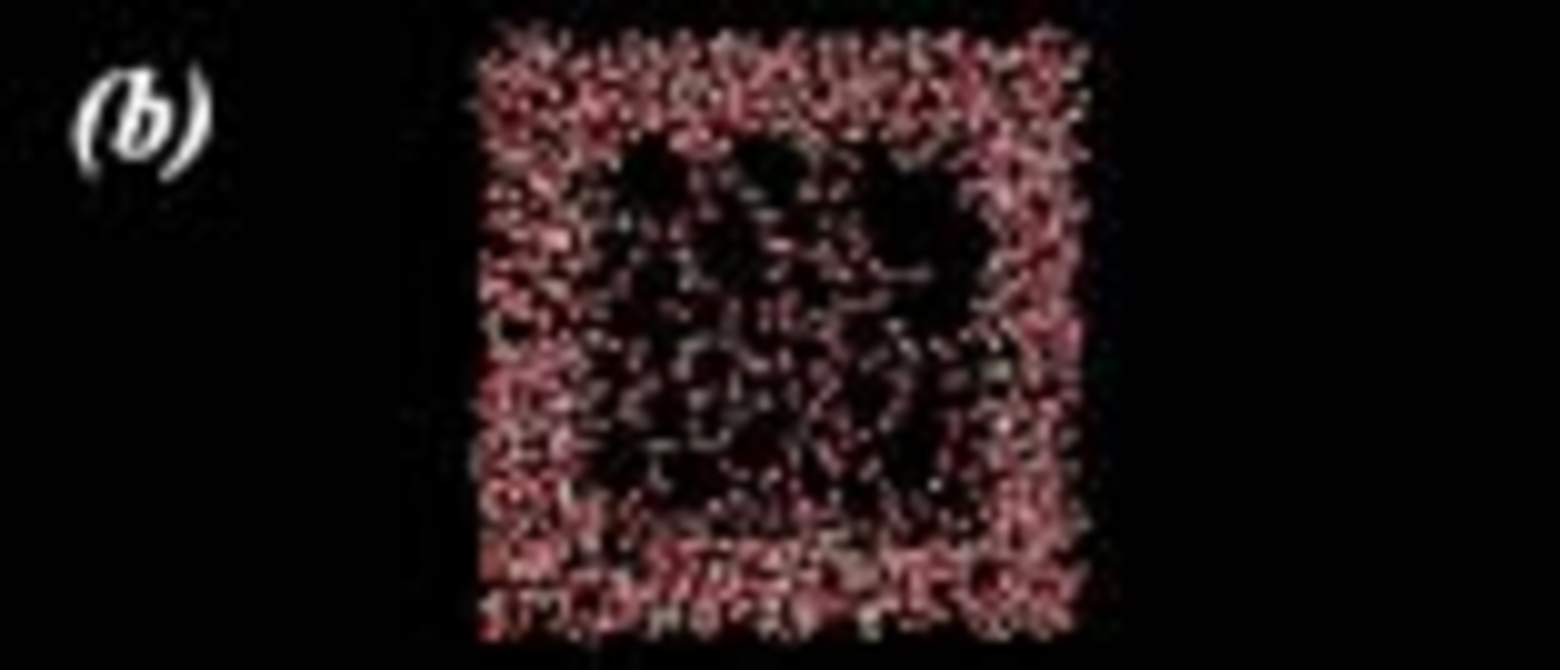}
} \centering{
    \includegraphics[width=3.0in]{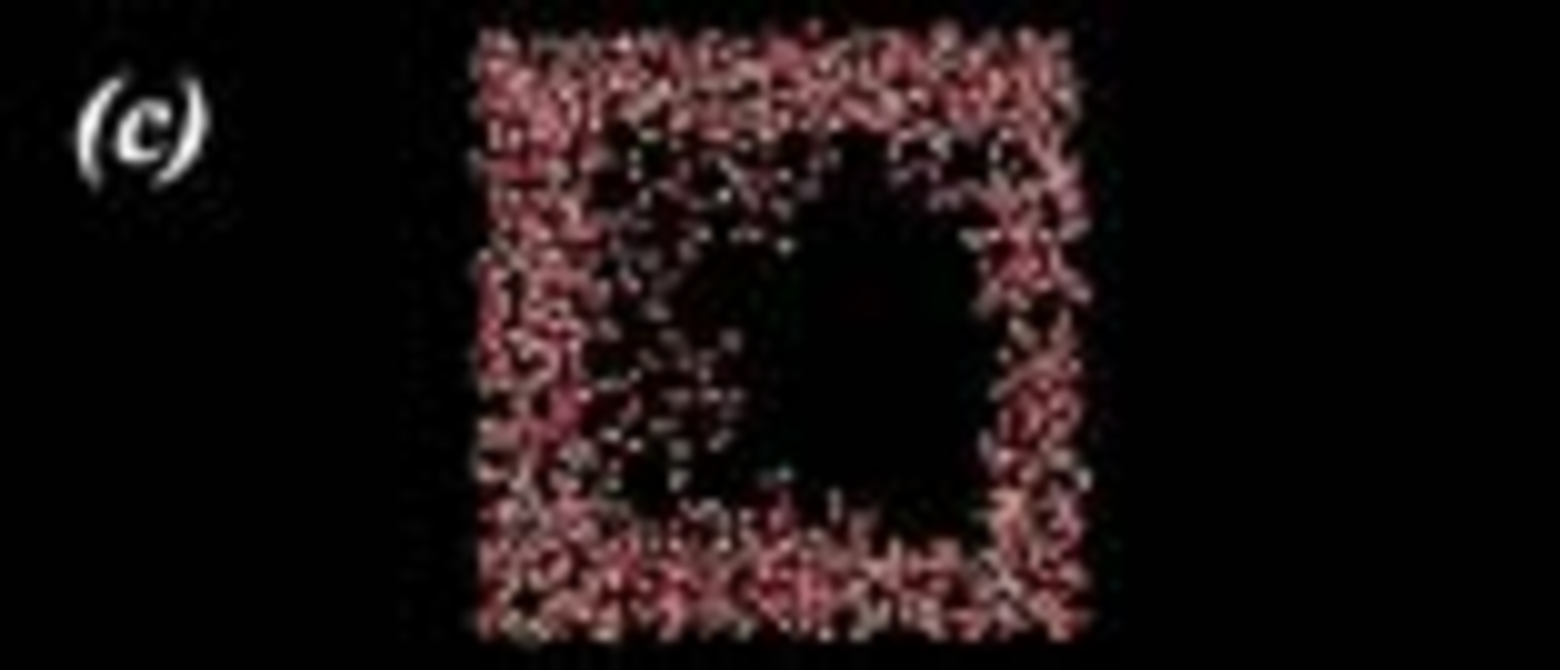}
    \includegraphics[width=3.0in]{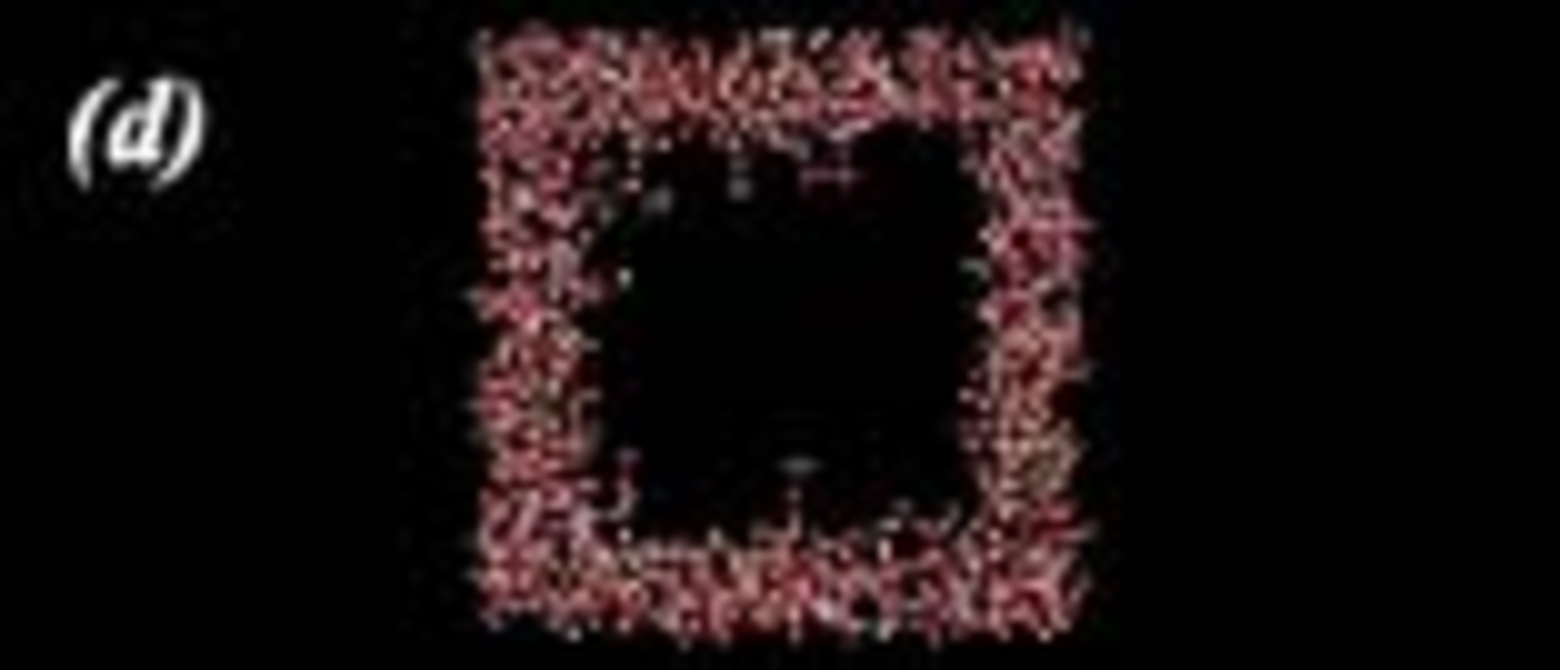}
} \caption{(Color online) Sequence of time-separated snapshots
showing capillary evaporation between hydrophobic plates at
$P=0$~GPa and $d=0.6$~nm. Times correspond to (a) $18$, (b) $40$,
(c)  $135$, and (d) $239$~ps. We only show water molecules in the
slab corresponding to the space between plates (i.e., the plates
themselves have been omitted for clarity) . The square cavity
evident at $239$~ps corresponds to the whole area of the plates.}
\label{dewettSequence}
\end{figure}

 A sequence of snapshots showing a typical dewetting
process is given in Fig. \ref{dewettSequence}. We observe that the
density rapidly becomes rarified on a time scale of $\approx
10$~ps, and then the solvent retreats on a somewhat longer time
scale ($\approx 100$~ps). At $P=0$~GPa and $d \leq 0.6$~nm, the
vapor cavity is limited to the area of the
 plates, i.e. the water surrounding the plates remains in the
liquid state [Fig. \ref{dewettSequence}(d)]. However, at the lower
pressure $P=-0.05$~GPa and $d \leq 1.2$~nm, the bubble originally
formed between the plates expands to the bulk liquid, inducing
cavitation of the whole system. In other words, a heterogeneous
nucleation event occurs induced by the plates. For $P \leq
-0.1$~GPa and $d\leq 1.6$~nm, we observe a simultaneous cavitation
inside and outside of the confined space indicating that, for the
present water model, the liquid phase is unstable for $P \leq
-0.1$~GPa. We note that when capillary evaporation occurs, no
water molecule is observed between the plates. Thus, practically
speaking, what we call `vapor phase' is here a vacuum.

Observations consistent with this picture have been reported in
measurements of the force between hydrophobically-coated mica
surfaces immersed in water \cite{claeson}. Specifically, vapor
cavities formed spontaneously when mica surfaces coated with a
double-chain cationic fluorocarbon surfactant approached to
distances between $1$ and $4$~nm \cite{claeson}. However, cavities
only formed after separation from contact when the mica surfaces
were coated with a double-chain cationic hydrocarbon surfactant
\cite{claeson}. Both the separation at which cavities formed and
the magnitude of the force between the hydrophobic surfaces were
found to be very sensitive to the specific substance used to coat
the mica surface \cite{pashley,claeson}. The measured width of the
dewetting layer formed at an extended paraffin surface in water
($\approx 1.5$~nm) is also consistent with the present
calculations \cite{jensen1}.

Our results on capillary evaporation are in agreement with
previous grand canonical ensemble Monte Carlo simulations of water
using the SPC model \cite{bratko}. That work shows that capillary
evaporation between smooth planar hydrophobic walls occurs at
walls separations $D \approx 1.27$~nm and $D \approx 0.9$~nm at
$P=0$~GPa and $P \approx 0.1$~GPa, respectively. These distances
correspond, in our case, roughly to $d \approx 0.79$~nm and $d
\approx 0.42$~nm (see Fig.~\ref{PhaseDiag-d-P}).

\begin{figure}
\includegraphics[width=3.375 in]{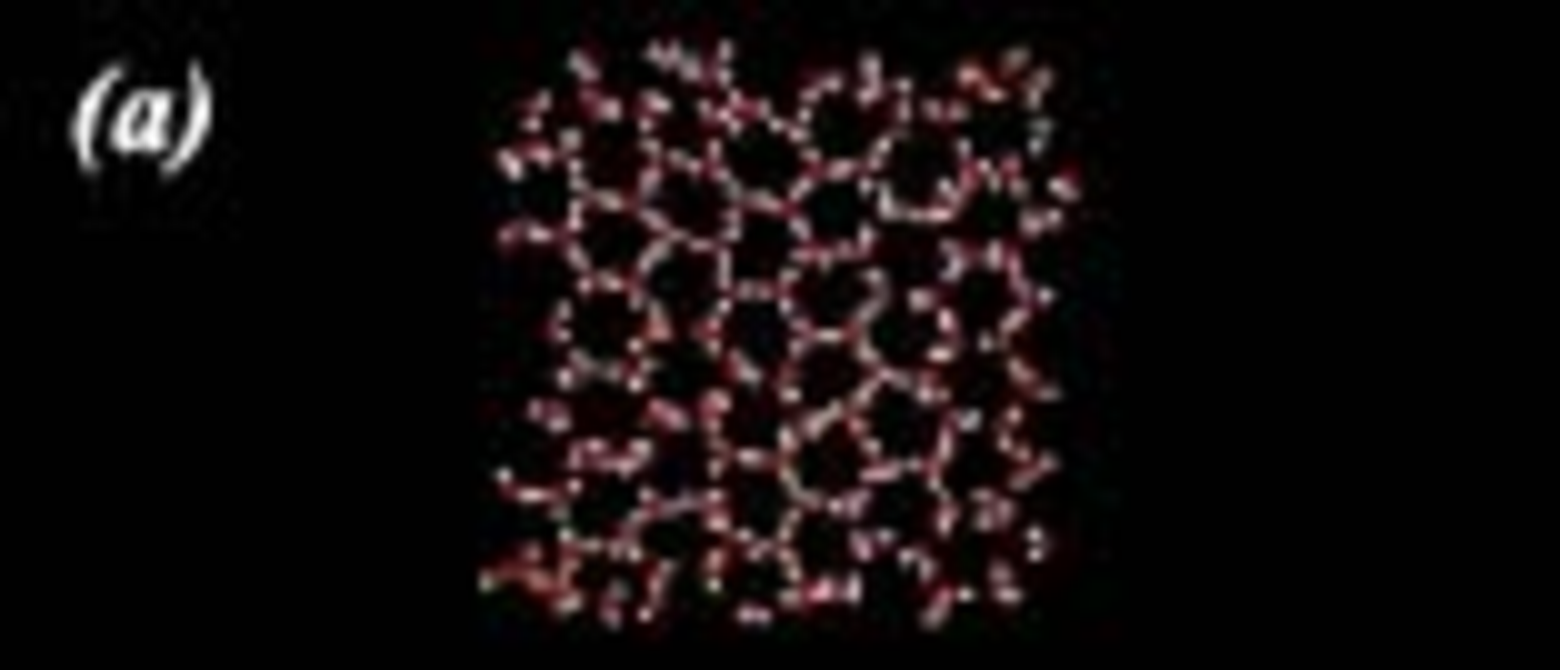}
\includegraphics[width=3.375 in]{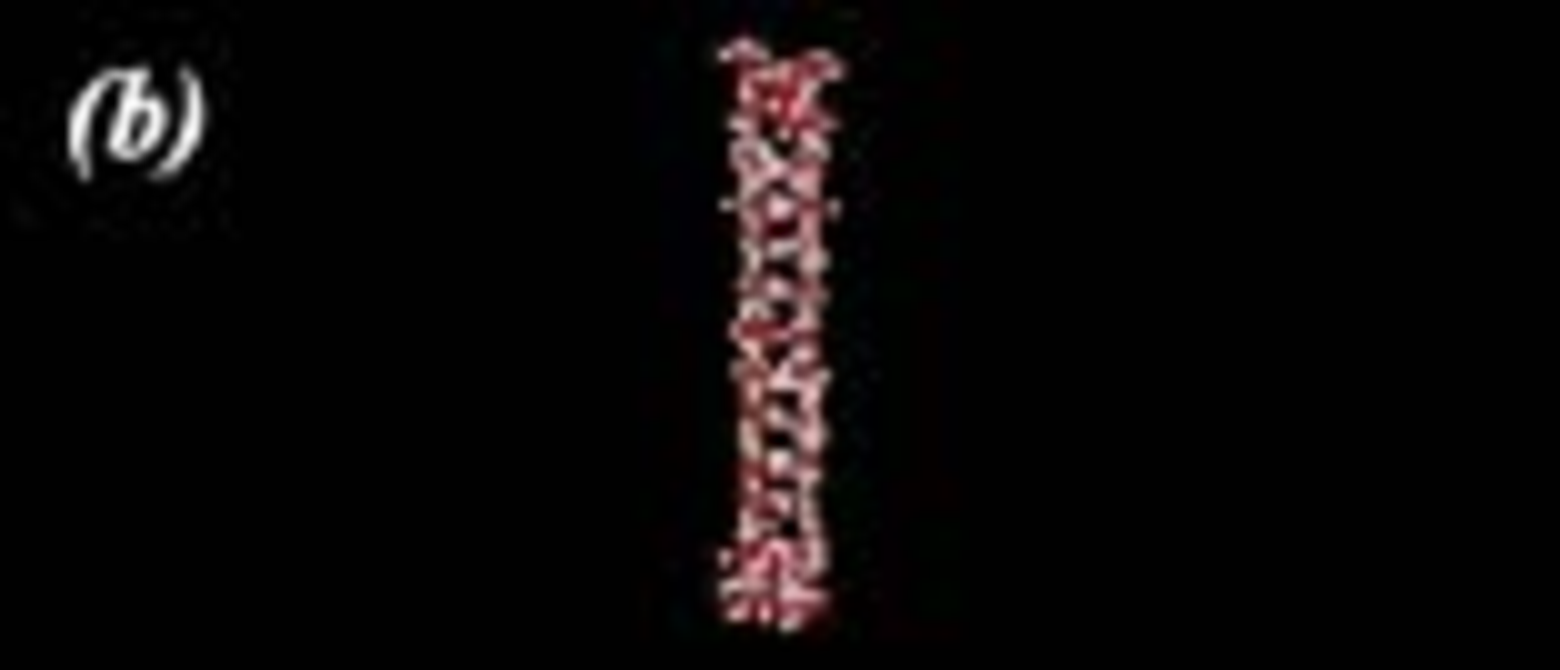}
\caption{(Color online) (a) Front and (b) lateral snapshots of the
bilayer ice formed between hydrophobic plates at $P=0.2$~GPa and
$d=0.6$~nm.}
 \label{iceSnapshot}
\end{figure}

The presence of a crystal at high $P$ has not been discussed so
far in the context of capillary evaporation. However, there is
considerable evidence from MD simulations showing that confined
water may crystallize at high $P$ (see e.g. \cite{zangiReview}).
Koga {\it et al.} found crystallization in MD simulations of the
TIP4P model at $T \approx 230-300$~K and $P=0.5-1.0$~GPa
\cite{KogaIcePrl,tanaka2005,kogaNat}. The ice structure that we
find is similar to that found in \cite{KogaIcePrl,kogaNat} and
resembles none of the structures of the existing ice phases in
bulk water, nor those found in metallic or hydrophobic pores
\cite{IceNotSeen1,IceNotSeen2,IceNotSeen3,IceNotSeen4}.
 Figure \ref{iceSnapshot}(a), a top view of the bilayer ice,
clearly shows a hexagonal lattice, while Fig. \ref{iceSnapshot}(b)
suggests that each layer is almost flat, and that the two layers
are in registry.

\begin{figure}
\centering{
    \includegraphics[width=2.7 in]{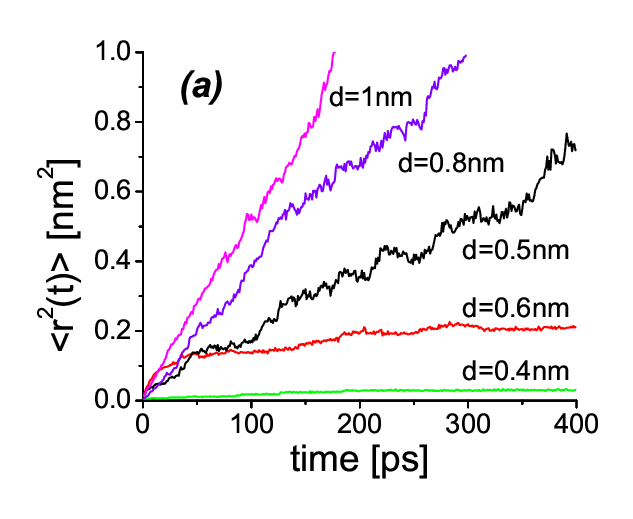}
    \includegraphics[width=2.7 in]{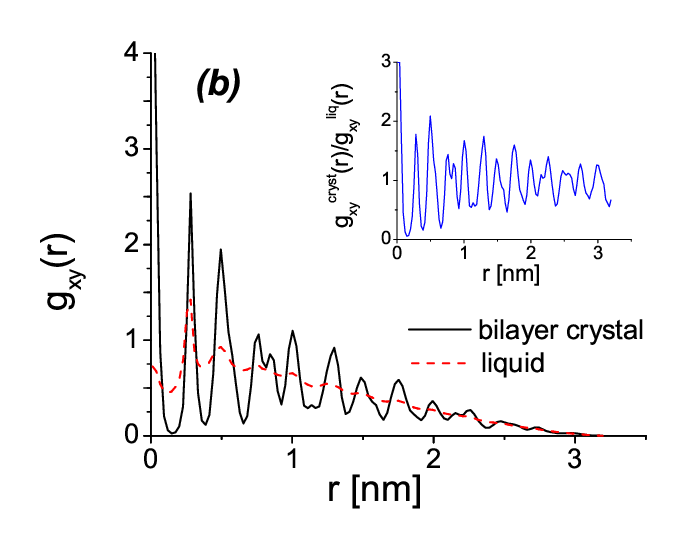}
    }
\centering{
    \includegraphics[width=2.7 in]{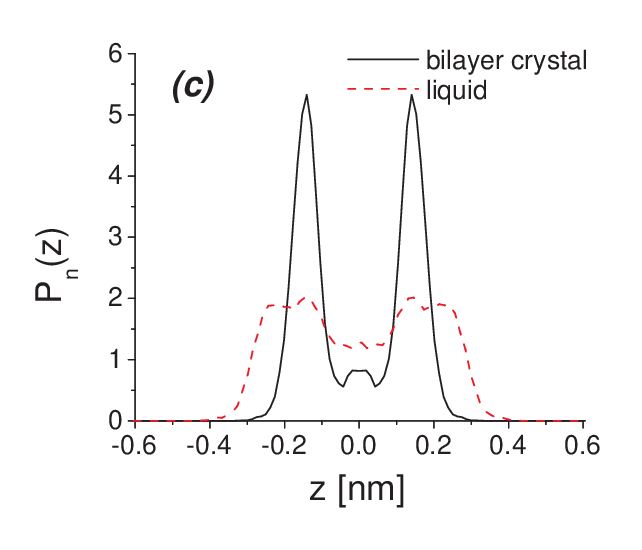}
}
 \caption{(Color online) (a) Lateral mean square displacement (MSD)
 as a function of time at $P=0.2$ GPa and different
plate-plate separations $d$. Hydrophobic plates. At $d=0.6$~nm the
confined water crystallizes and the MSD for long times is smaller
than for $d=0.5,~0.8$ and $1.0$ nm (where the system is in the
liquid phase). For comparison, we also show the results for
$d=0.4$~nm where water molecules are trapped between the plate
atoms due to the small value of $d$. (b) Radial distribution
function (RDF) parallel to the plates, $g_{xy}(r)$, for the
crystal ($d=0.6$~nm) and liquid ($d=0.8$~nm) phases. Inset: ratio
of the crystal RDF to that of the liquid. (c) Probability density
function associated with the distribution of water molecules
between the plates ($\int P_n(z)~dz=1$).}
 \label{msdIce}
\end{figure}

To confirm that confined water crystallizes at high $P$ and $d
\approx 0.6$~nm, we calculate the mean square displacement (MSD)
parallel to the plates at $P=0.2$ GPa for different values of $d$.
Figure~\ref{msdIce}(a) shows the MSD for a $400$~ps simulation
averaging over molecules (we do not average over starting times).
For $d =0.4$~nm, we find that MSD $\approx 0$ consistent with the
earlier observation that molecules are trapped between the plate
atoms. For $d=0.5,~0.8$, and $1.0$~nm the MSD increases
monotonically with time and, for a fixed time, with $d$. However,
the MSD for $d=0.6$~nm shows a fast increase for $t \lesssim
50$~ps and then approaches an asymptotic value of $\approx
0.2$~nm. Snapshots of the system during the first $50$~ps show
that the molecules in the confined space reorganize to form the
crystal. For $50$~ps~$\lesssim t \lesssim 300$ps, the resulting
ice shows defects that
 disappear with time. The transformation is almost over for $t\approx 300$~ps.
In the absence of crystallization, one would expect the MSD at
long times for $d=0.6$~nm to fall between that corresponding to
$d=0.5$~nm and $d=0.8$~nm. Instead, the MSD curve for $d=0.6$~nm
shows a plateau, as for $d=0.4$~nm, after about $250$~ps,
consistent with the view that the system crystallizes.

The crystallization of confined water is also confirmed by
structural properties such as the radial distribution function
(RDF) parallel to the plates, $g_{xy}(r)$,
 and the probability density function to find a molecule located at $z$ between
 the plates, $P_n(z)$.
Figure~\ref{msdIce}(b) shows $g_{xy}(r)$ for the ice ($d=0.6$~nm)
 and for a bilayer liquid ($d=0.8$~nm). The oscillations in $g_{xy}(r)$
for $d=0.6$~nm are a clear sign of crystallization. Such
oscillations are not present in the liquid phase. We note that the
slow decay of $g_{xy}(r)$ for large $r$ is due to the finite size
of the plates. This finite size effect can be removed by taking
the ratio of $g_{xy}(r)$ for the crystal to that of the liquid
(see inset). The persistence of oscillations is then clear.
Figure~\ref{msdIce}(c) shows the $P_n(z)$ for the ice ($d=0.6$~nm)
and for a comparable liquid ($d=0.8$~nm). While in both cases
water molecules form two layers \cite{footnoteLayer}, the peaks of
$P_n(z)$ are much larger at $d=0.6$~nm than at $d=0.8$~nm.
$P_n(z)$ was calculated from a histogram of water molecules in
$78$ slices into which the distance between plates was divided. It
is defined so that $\int P_n(z)~dz = 1$.

\begin{figure}
\centering{
    \includegraphics[width=2.8 in]{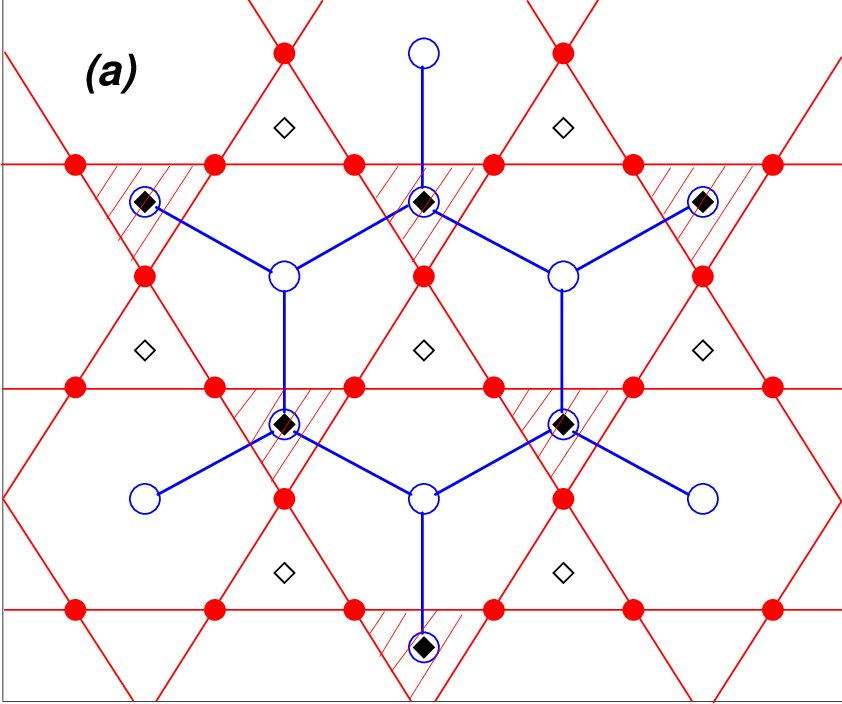}
    }
\centering{
    \includegraphics[width=2.8 in]{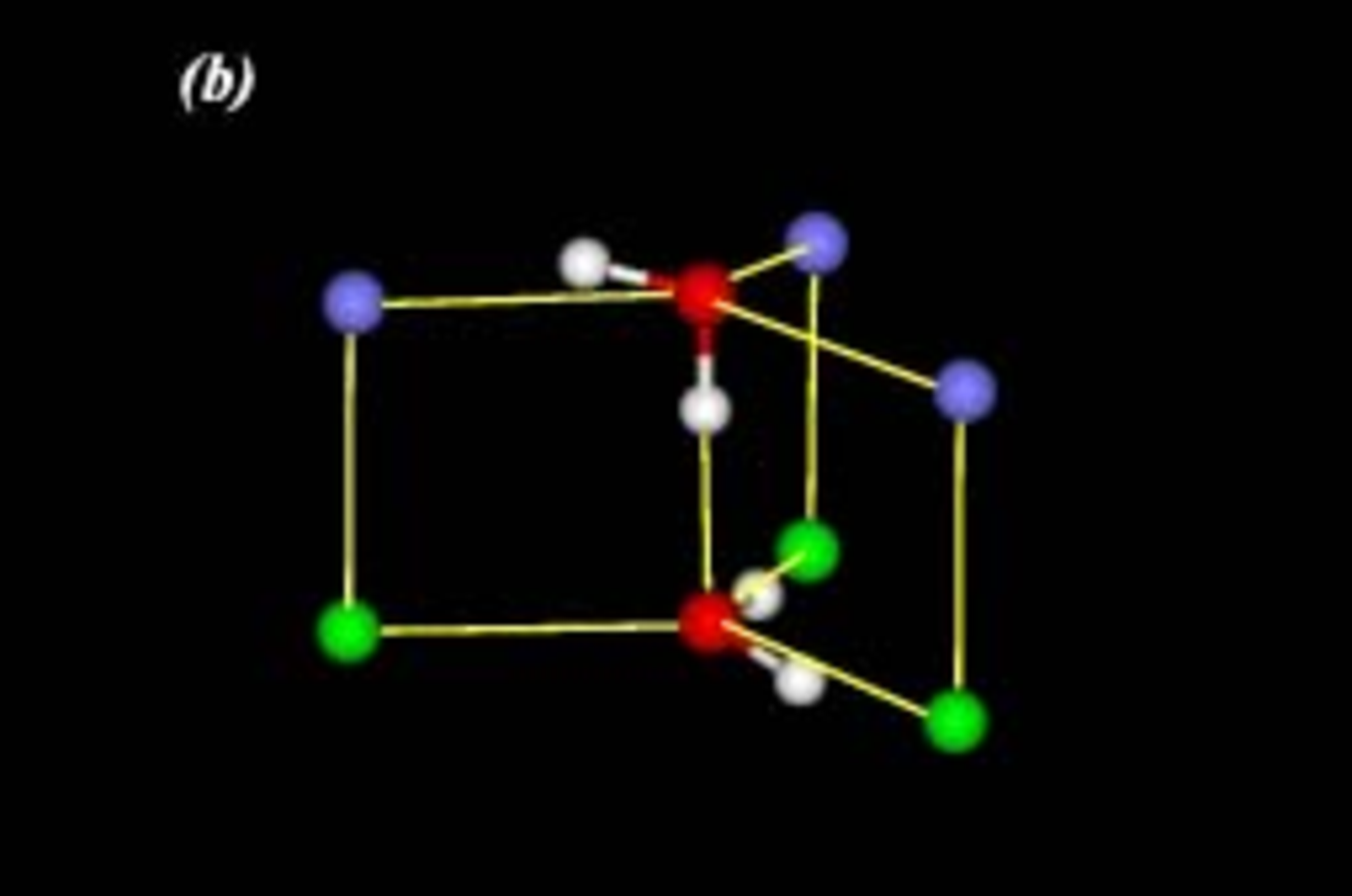}
    \includegraphics[width=2.8 in]{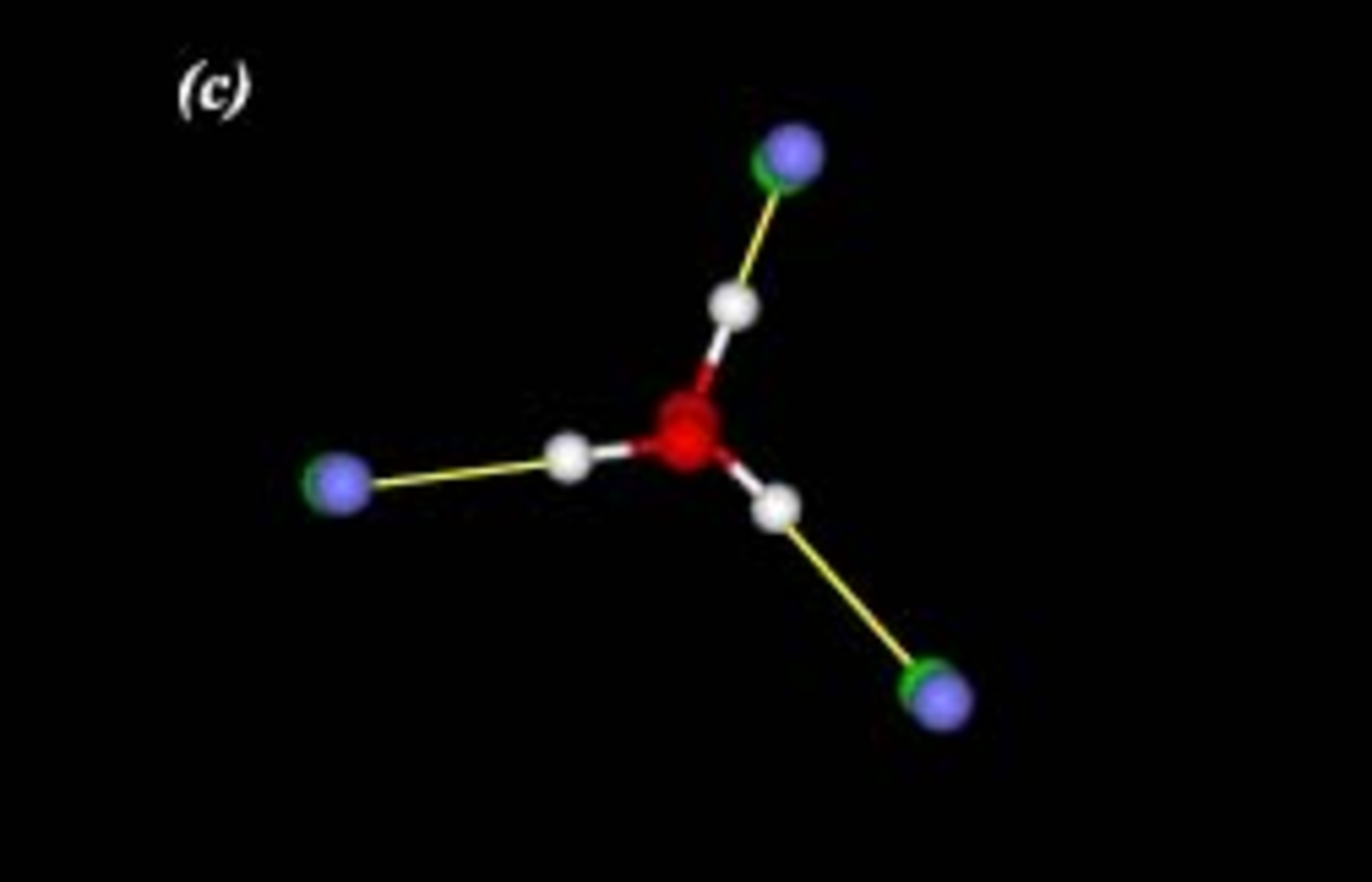}
} \caption{(Color online) (a) Schematic diagram showing the
location of water molecules in the bilayer ice phase relative to
the position of the hydrophobic plate surface atoms. Red filled
circles represent `oxygen' atoms belonging to a single plane on
the plate surface. The next layers of `oxygen' atoms, above and
below this plane, are represented by empty and filled black
diamonds, respectively. The atoms at the plate surface are
arranged in tetrahedra pointing either into or out of the page of
the figure [{\it cf.} Fig.~\ref{SnapshotWalls}~(d),(e)].
Tetrahedra pointing into the page correspond to the cross-hatched
triangles centered on the filled diamonds, while those pointing
out of the page correspond to triangles centered on the empty
diamonds. Water oxygen atoms are represented by empty circles; a
typical hexagon is indicated by blue lines. Lateral (b) and top
(b) view of molecules belonging to the bilayer ice, showing the
hydrogen bonds. We show only two water molecules with the
corresponding hydrogen atoms (red and white spheres). These
molecules belong to different layers of the ice and their
corresponding nearest neighbors are represented by blue and green
spheres. Yellow lines are a guide to the eye showing the lattice
characterizing the bilayer ice. } \label{schemeIce}
\end{figure}

 We note
that no spontaneous crystallization in bulk or confined water has
been reported in MD simulations using the SPC/E model.
Crystallization using the TIP4P \cite{ohmineNat} and TIP5P models
has been obtained \cite{masako}. In our case, we note that the
location of water molecules belonging to each of the two ice
layers is highly correlated with the plate structure, suggesting
that the crystal we obtain is templated by the plate. Such
substrate-templated crystallization of interfacial water has been
observed on {\it hydrophilic} surfaces \cite{ocho}.
 Figure~\ref{schemeIce}(a) is
a schematic diagram showing both the hexagons formed by the water
oxygen atoms and the atoms at the plate surface. The atoms at the
plate surface are arranged in tetrahedra pointing either into or
out of the page of the figure. These two kinds of tetrahedra
alternate, forming rings composed of six tetrahedra [see also
Fig.~\ref{SnapshotWalls}(a)]. Two rules determine the structure of
a given hexagon of the ice layers: (i) a water oxygen atom is
located on top of each tetrahedron pointing into the page of the
figure; and (ii) a water oxygen atom is located on top of the
center of the hexagonal rings formed by the tetrahedra at the
plate surface. As a result, the ice layers are formed by perfect
hexagons with an $OO$ distance of $0.29$~nm. Interestingly, this
is the same distance separating the two ice layers [see
Fig.~\ref{msdIce}(c)]. Thus, the $OO$ distance between nearest and
next nearest neighbors are $0.29$~nm and $\sqrt{3} \times
0.29$~nm~$ = 0.49$~nm, respectively. These values are in agreement
with the location of the first two maxima of $g_{xy}(r)$ [see
Fig.~\ref{msdIce}(b)] which are located at $\approx 0.29$~nm and
$\approx 0.51$~nm, respectively.

The bilayer ice is fully hydrogen-bonded and each molecule has
four HBs, as in ordinary ice. HBs occur either between molecules
in the same ice layer or between both layers (i.e., there are no
HB pointing toward the plates). The two water layers are linked by
HBs where a molecule of one layer shares one of its $H$ atoms with
the nearest oxygen atom of the other layer [see
Figure~\ref{schemeIce}(b) and (c)].  We note that the $HOH$ angle
in the SPC/E model is $109.47^o$, i.e., the tetrahedral angle.
However, in the bilayer ice structure, the $OOO$ angles (formed
between three nearest neighbors) are either $90^o$ or $120^o$.
Therefore, in general, the $OO$ directions (between nearest
neighbors) differ slightly from the $OH$ directions of a given
water molecule.

In refs. \cite{mashl,kogaTube}, it was found that TIP4P water
within carbon nanotubes can crystallize forming a tube of square,
pentagonal, or hexagonal ice, depending on the nanotube radius.
The structure of the carbon nanotube hexagonal ice seems to be the
same as that found in the present work. Interestingly, the same
requirements [that the confinement geometry allows: (i)
next-nearest oxygen atoms to be located roughly at the bulk O-O
distance ($\approx 0.28$~nm) \cite{kogaTube}; and (ii) water
molecules to have four hydrogen bonds] are necessary for ice
formation in our case and in the carbon nanotubes.

Simulations of water between crystalline infinite walls using the
TIP5P model \cite{zangi1,zangi2,zangiReview}, and between smooth
infinite walls using the TIP5P \cite{pradeep} and TIP4P models
\cite{tanaka2005} also show the formation of crystals at
$T=300$~K. Both monolayer and tri-layer ices have been reported
\cite{zangi2,pradeep}. As indicated in Fig. \ref{PhaseDiag-d-P},
our simulations at $P=0.2$~GPa show no sign of monolayer ice at
small $d$ nor of crystallization to other solids at larger $d$
(corresponding to $n$-layer crystals, $n \geq 3$). Small
variations in $d$ ($\lesssim 0.1$~nm) were found to have a
profound effect on the appearance and disappearance of confined
ices \cite{zangi1,zangi2,tanaka2005}. In Fig.~\ref{PhaseDiag-d-P},
the values of $d$ simulated at a given $P$ are separated by
$\Delta d \geq 0.1$~nm. Thus, it is possible (but unlikely) that
these ices can be found for confined SPC/E at $T \approx 300$~K
when using a smaller $\Delta d$.

Figure \ref{PhaseDiag-d-P} also shows a schematic confined phase
diagram superimposed on the simulation data points. The bilayer
crystal and liquid phases are indicated, together with an estimate
of the dewetting transition line. At $T=300$~K, we find that the
dewetting transition line extends down to $d=0.5$~nm, suggesting
that it merges with the region (defined by $d \lesssim 0.4$~nm)
for which molecules gets trapped between the plate atoms. It is
interesting to consider whether for a higher $T$ the dewetting
transition line might end in a critical point.
%In other words, if the bulk water gas-liquid critical point is
%located at $(P_c,T_c)$, then at $T \approx T_c$ one could expect
%to find a critical point in the $d-P$ plane at $P \approx P_c$ and
%large $d$. After all, for large $d$, most of the water confined
%between the plates will behave like bulk water.
The relationship between the dewetting locus and water's ordinary
vapor-liquid critical point will be investigated in future studies
at higher temperatures. It is also possible that the bilayer
crystal region in Fig.~\ref{PhaseDiag-d-P} expands down to the
dewetting transition line (see red dotted line in the figure). In
this case, there would be a triple point where the vapor phase
coexists with the liquid and crystal phases. Our simulation at
$P=0.05$~GPa and $d=0.6$~nm does not show the presence of a
crystalline structure. Instead, we observe a competition between a
disordered (liquid) structure and bubble formation. At $600$~ps a
bubble occupies $25\%$ of the confined space, and at $1$ ns we
find that $90\%$ of the confined space is dewetted. With the size
of the plate area
 that we simulate, it is not possible to observe simultaneously
 the vapor, liquid, and crystal phases [only approximately five
hexagons per side can be clearly observed in
Fig.~\ref{iceSnapshot}(a)] . A precise investigation of the
existence (or lack thereof) of a triple point requires larger
plate surface areas than we have used in this work.

\subsection{Water structure: effect of pressure}
\label{vconfSect}

To study the effects of pressure on the structure of water we fix
$d=1.6$~nm. This separation is large enough so that water
properties in the middle region between the plates are essentially
those of bulk water \cite{lee1}.

\begin{figure}
\includegraphics[width=3.375 in]{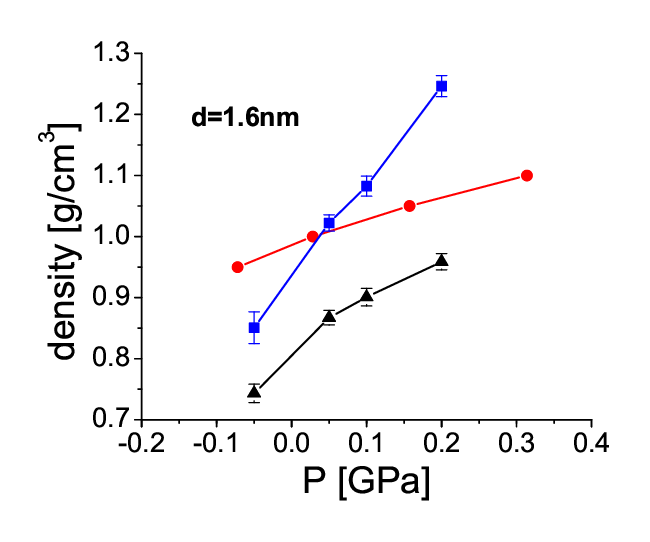}
\caption{(Color online)  Average density $\rho(P)$
 for different values of $P$ in the confined space between hydrophobic plates, and in  bulk water.
 Circles correspond to the values of $\rho(P)$ for bulk water
  taken from ref. {\protect \cite{francislong}}. For confined
  water, we show two estimates of $\rho(P)$ corresponding to
  different definitions of the confined volume.
   Squares and triangles correspond to $\rho(P)$ calculations
 when using the `effective' and `formal' definition of volume, respectively (see
 Sect.~\ref{vconfSect} for details).
}
 \label{nO-tPhi0.0}
\end{figure}

\begin{figure}
\centering{
\includegraphics[width=2.8 in]{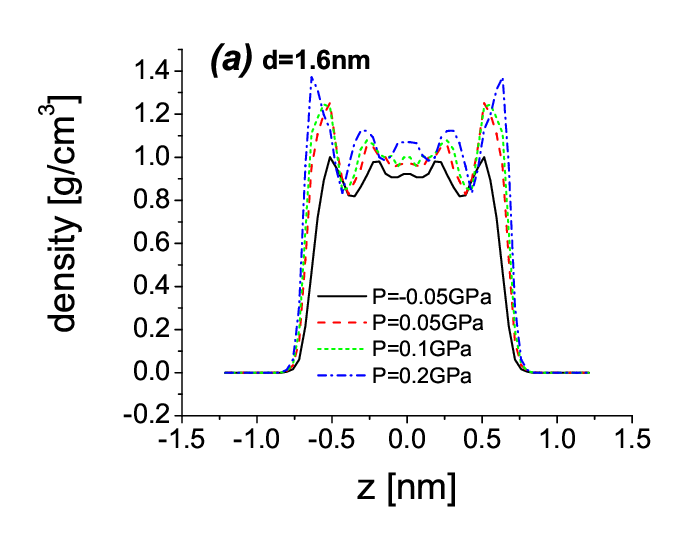}
\includegraphics[width=2.8 in]{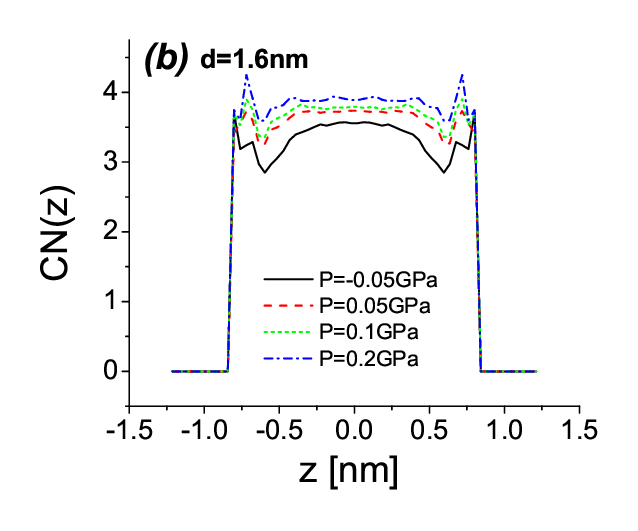}
} \centering{
\includegraphics[width=2.8 in]{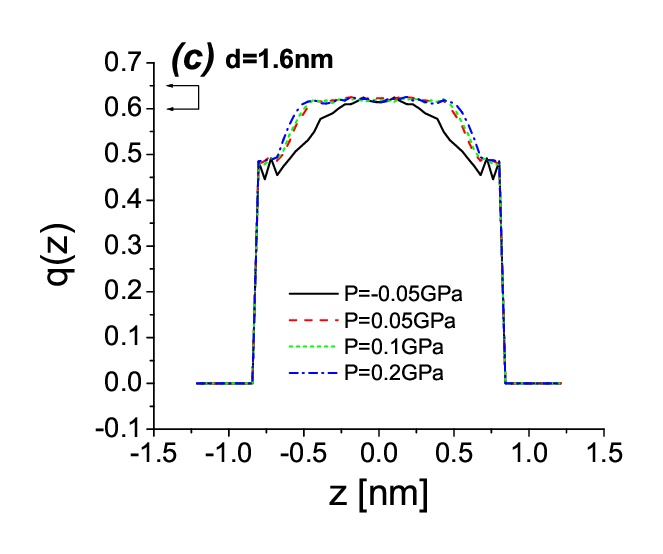}
}
 \caption{(Color online) (a) Density profile, $\rho_{slab}(z)$
 (i.e., the local density in a slab of width $0.0411$~nm, parallel to the
 plates, located at $z$); (b) average coordination number $CN(z)$; and (c) order parameter $q(z)$.
% in slabs of width $0.0411$~nm parallel to the plates, located at $z$.
Arrows in (c) indicate the range of $q$ values for bulk water at
$T=300$~K and $0.85$~g/cm$^3$~$\leq \rho \leq 1.15$g/cm$^3$
{\protect \cite{errington}}. Hydrophobic plates.}
 \label{X-zPhi0.0}
\end{figure}

%Figure \ref{nO-tPhi0.0}(a) shows the typical time dependence of the
%total number of molecules between the plates $n_O(t)$. The
%corresponding average density at each pressure, $\rho(P)$, is
%shown in Fig. \ref{nO-tPhi0.0}(b).
Figure \ref{nO-tPhi0.0} shows the average water density between
the plates at each pressure, $\rho(P)$. For comparison, we show
also the density of bulk water reported in \cite{francislong}. We
define $\rho(P)= \langle n(t)\rangle \times m_{H2O}/V_{conf}$
where $m_{H2O}$ is the mass of a water molecule and  $\langle n(t)
\rangle$ is the average number of molecules in the confined space
between the walls. When dealing with confined systems, defining
 the accessible volume for the confined liquid is not unique. We estimate
$V_{conf}$ in two ways: by the formal dimensions of the confined
space, based on the defined $d$, i.e., $V_{conf}=1.6 \times 3.217
\times 3.215$ nm$^3$; and by an effective available volume
$V_{conf}= d_{eff} \times 3.217 \times 3.215$ nm$^3$ where
$d_{eff}$ is an effective plate-plate distance. If $P_n(z)$ is the
probability density to find an $O$ atom in a slab parallel to the
plates located at $z$ ($\int P_n(z)~dz =1$), then $d_{eff}$ is
defined such that $P_n(\pm d_{eff}/2)=0.5$ [see
Fig.~\ref{X-zPhi0.0}(a)]. The first approach to the calculation of
$V_{conf}$ underestimates $\rho(P)$ because the formal confined
volume extends practically to the plate surface (i.e., the plane
where the $H$ atoms would be located), and thus includes space not
accessible in practice to water molecules. The second (effective)
definition of $V_{conf}$ overestimates $\rho (P)$ because it
leaves out from the calculation a portion of volume that is in
fact accessible to water molecules.

$\rho(P)$ in the confined space is always smaller than the bulk
density for $P \lesssim 0.06$~GPa, independently of the method
used to calculate $V_{conf}$. For $P > 0.06$~GPa the two estimates
of $V_{conf}$ produce drastically different results. One expects
that at high $P$ the water density in the confined space
approaches that of the bulk water. It can be seen from
Fig.~\ref{nO-tPhi0.0} that the formal definition of $V_{conf}$,
using the $d$-value defined a priori, leads to a more reasonable
high-$P$ trend. For all cases, one finds that the slope of
$\rho(P)$ is always greater than that of bulk water indicating
greater compressibility between the plates. To confirm this, we
interpolate $\rho(P)$ in Fig.~\ref{nO-tPhi0.0} with a second order
polynomial, and calculate the corresponding derivative with
respect to P. The resulting values for the compressibility,
$\kappa_T$, at $P=0$~GPa and $T=300$~K are:
$\kappa_T~0.47$~GPa$^{-1}$, for bulk water; and
$\kappa_T~1.47$~GPa$^{-1}$ (formal volume) and $\kappa_T~
1.70$~GPa$^{-1}$ (effective volume), for confined water. The
experimental value of the compressibility at these conditions is
$\kappa_T~0.45$~GPa$^{-1}$ \cite{compressExp}, in good agreement
with the simulations.

Figure \ref{X-zPhi0.0}(a) shows the density profile
$\rho_{slab}(z)$, i.e. the local density in a slab parallel to the
plates, located at $z$. It was calculated using $39$ slabs of
width $0.0411$~nm, with an area equal to the plates area.
$\rho_{slab}(z)$ resembles the number density at normal $T$ and
$P$ reported in Ref. \cite{lee1}. Figure \ref{X-zPhi0.0}(a)
indicates that the main effect of increasing $P$ is to push water
molecules toward the plates. Accordingly, as $P$ increases,
$\rho_{slab}(z)$ increases for $z \approx 0$ and the maxima at $z
\pm 0.51$~nm for $P=-0.05$~GPa shift to $z \pm 0.63$~nm at
$P=0.2$~GPa. Moreover, the maxima and minima of $\rho_{slab}(z)$
are more pronounced at $P=0.2$~GPa than at $P=-0.05$~GPa, as
molecules increasingly sample `harder' plates at high $P$. We
obtain the same conclusions when looking at the probability
density function for the H atoms.

To analyze the local packing and hydrogen bonding, we calculate
the average coordination number $CN(z)$ and the order parameter
$q(z)$ used to characterize the local tetrahedral order in water
\cite{errington,ourGlass}. Averages are taken over slabs of width
$0.0411$~nm centered at different $z$. We define $CN(z)$ as the
number of neighbor oxygens within a distance $\leq 0.32$~nm (the
first minimum of the radial distribution function at
$\rho=0.984$~g/cm$^3$ and $T=284$~K \cite{francesco}) from a
central $O$ atom. $q(z)$ is defined as \cite{errington}
\begin{eqnarray}
q \equiv 1 - \frac{3}{8} \sum_{j=1}^3 \sum_{k=j+1}^4 \left( \cos
\psi_{jk} + \frac{1}{3}    \right)^2 \label{Qdef}
\end{eqnarray}
where $\psi_{jk}$ is the angle formed by the lines joining the
oxygen atom of a given molecule and those of nearest neighbors $j$
and $k$ ($\leq 4$). In this work, we also include the oxygen ($O$)
atoms in the plates when considering the $O$ nearest neighbors $j$
and $k$.

Figure \ref{X-zPhi0.0}(b) shows $CN(z)$ for different values of
$P$. At all $P$, we find that $CN(Z)$ for $z \approx 0$ is close
to $4$ as is the case for bulk water. The slight increase of
$CN(z)$ with $P$ at $z=0$ can be due to the fact that we are using
a fixed cutoff of $0.32$~nm to calculate $CN$ (while the first
minimum of the radial distribution function depends slightly on
$P$). We observe that $CN(Z)$ decreases as one approaches the
plates. However, right next to the plates we find sharp maxima.
Moreover, such maxima increase with $P$ as water molecules are
pushed toward the plates [see Fig.~\ref{X-zPhi0.0}(a)]. When the
$O$ atoms in the plates are not considered in the calculation of
$CN(z)$, we find a distribution similar to that of \cite{lee1}. In
this case, $CN(z) \approx 2.5$ next to the plates, indicating that
molecules in contact with the surfaces lose on average $1.5-2$
{\it water} neighbors.

Figure \ref{X-zPhi0.0}(c) shows $q(z)$ for different values of
$P$. At all $P$, we find that $q(z) \approx 0.62$ at $z=0$, close
to the value $0.60-0.65$ for bulk water at $0.85 \leq \rho \leq
1.15$~g/cm$^3$ and $T=300$~K \cite{errington} (indicated in the
figure). The peaks of $CN(z)$ next to the plates are accompanied
by the rapid decrease of $q(z)$ approaching the interface. At the
plates, $q(z)\approx 0.5$, and this value does not change with
$P$, that is, the local tetrahedrality is not apparently affected
by $P$ for molecules in contact with the surfaces.

\begin{figure}
\includegraphics[width=3.375 in]{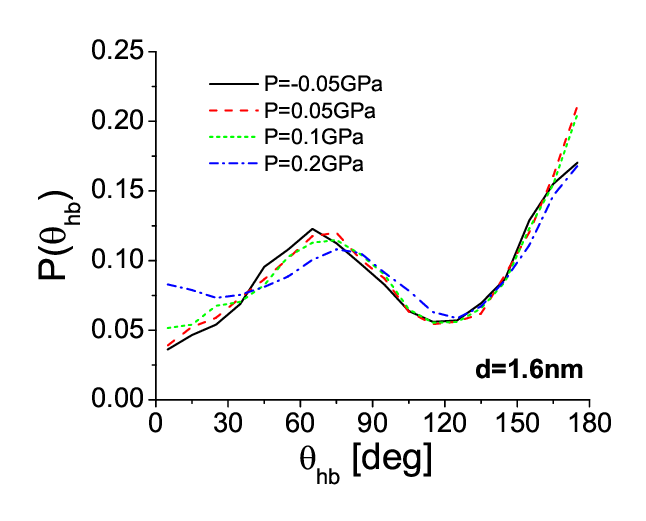}
\caption{(Color online) Normalized distribution $P(\theta_{hb})$
of angles $\theta_{hb}$ between the hydrogen bond vectors of water
molecules and the inward pointing normal to the plates. Average is
performed over molecules at a distance $\leq 0.2$~nm from the
plates. Hydrophobic plates.
%$P(\theta_{hb})$ shows maxima at $\theta_{hb}= 70^o$ and $180^o$
%indicating that in average water molecules have one HB vector
%pointing toward the nearest plate.
}
 \label{TitaHbPhi0.0}
\end{figure}

To study the orientational structure, we follow \cite{lee1} and
compute the distribution of angles $\theta_{hb}$ between the four
HB vectors of water molecules and the inward pointing normal to
the plates. Each molecule is associated with four HB vectors:
these four vectors point tetrahedrally outwards from the oxygen
Lennard-Jones site, and such that two of them join the $O$ site to
the same molecule's $H$ atoms. The normalized distribution
$P(\theta_{hb})$ in the proximity of the plates is shown in Fig.
\ref{TitaHbPhi0.0} for different values of $P$.  $P(\theta_{hb})$
shows maxima at $\theta_{hb}= 70^o$ and $180^o$ indicating a
preferred orientation for a water molecule with one HB vector
pointing toward the nearest plate. These results, and a
$P(\theta_{hb})$ calculated in slabs at different distances from
the plates (not shown), are in agreement with the results obtained
in ref. \cite{lee1} in constant volume MD simulations.
Interestingly, we find that when $P$ increases, despite changes in
$\rho_{slab}(z)$,
 $P(\theta_{hb})$ barely changes, as is evident in Fig.~\ref{TitaHbPhi0.0}.
  In particular, molecules next to the plates preserve their distinct
   orientational order under pressure.
 This is in agreement with the noted invariance of $q(z) \approx
0.5$ next to the surfaces for all $P$.

\begin{figure}
\includegraphics[width=3.375 in]{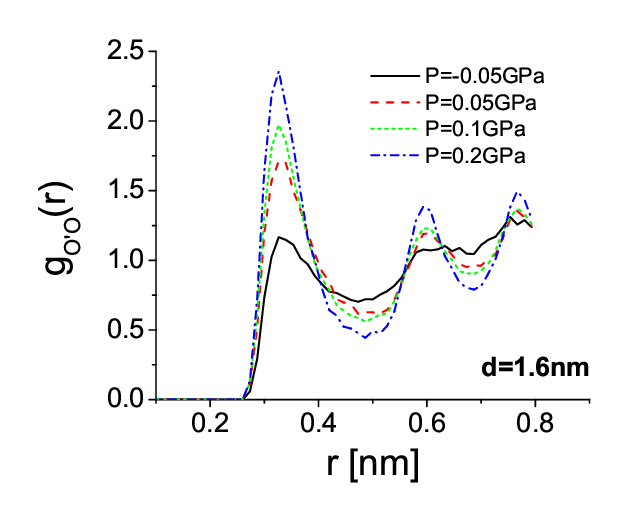}
\caption{(Color online) Radial distribution function for the water
molecule $O$ and plate surface $O$ atoms.  Hydrophobic plates.}
 \label{rdfPhi0.0}
\end{figure}

The radial distribution function (RDF) for the water molecule $O$
atom and the $O$ atom at the surface of the plate, $g_{O'O}(r)$,
is shown in Fig. \ref{rdfPhi0.0}. At all $P$, $g_{O'O}(r)$ has
maxima at $0.32$, $0.60$, and $0.77$ nm. For comparison, the first
maximum of the bulk water RDF at $\rho=0.984$~g/cm$^3$ and $T=
284.5$~K \cite{francesco} is at $0.28$~nm. Thus, water oxygen
atoms are located farther from plate surface oxygen atoms than
from other water molecule oxygen atoms, as expected for
hydrophobic surface atoms.
%Of course, this is due to the lack of polarity associated with the
%plate atoms.
The increase of $P$ does not shift the location of the maxima and
minima of $g_{O'O}(r)$, but changes their relative heights. As $P$
increases, more molecules are found in the confined space and, at
the same time, more molecules shift from interstitial to
coordination shells. Thus, the confined liquid becomes more
structured upon compression.

In summary, the effect of pressure on the structure of water
confined between hydrophobic plates is to enhance water structure,
pushing the water molecules toward the plates. The orientation of
water molecules next to the plates is not sensitive to
compression, as indicated by Fig. \ref{TitaHbPhi0.0}.

\section{Results: Hydrophilic plates}
\label{Hphilic}

Over the range of conditions investigated here ($T=300$~K,
$P=-0.05$, $0.05$, and $0.2$~GPa, and in each case $d=0.6$, $1.0$,
and $1.6$ nm), we find that the confined water remains in the
liquid state, and shows no sign of capillary evaporation or
crystallization. In agreement with the hydrophobic plate
simulations, we find that for $P\leq -0.1$~GPa, the whole system
cavitates; i.e., liquid water is unstable. Experimentally, this
would manifest itself as loss of cohesion and the appearance of a
macroscopic vapor phase; computationally, it becomes impossible to
maintain tensions in excess of $0.1$~GPa and the system volume
grows uncontrollably due to the appearance of the vapor phase.
However, even in this case we find no capillary evaporation
between hydrophilic plates. In other words, liquid water fills the
confined space, and the hydrophilic plates {\it induce wetting}
even when there is bulk cavitation around the plates.

\subsection{Water structure: effect of pressure}
\label{hphil-P}

\begin{figure}
\includegraphics[width=3.375 in]{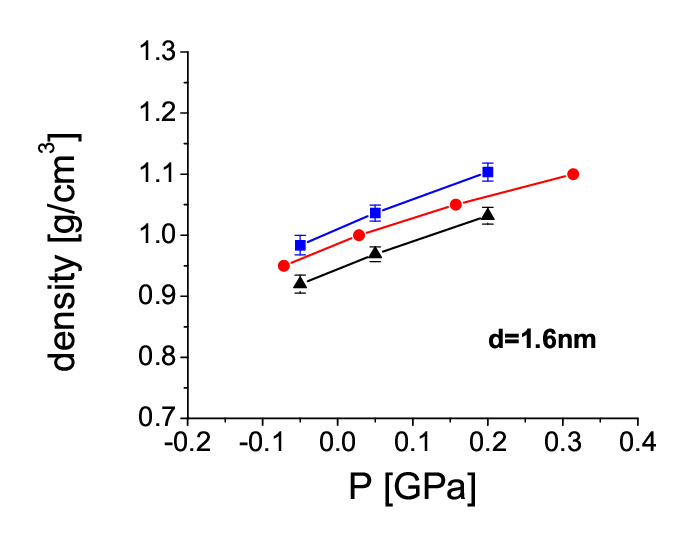}
\caption{(Color online) Density of water between hydrophilic
plates compared to that of bulk water. Symbols and explanation are
the same as in
 Fig.~\ref{nO-tPhi0.0}.}
 \label{nO-tPhi1.0}
\end{figure}

 As previously done for the case of the hydrophobic
plates, we chose a distance between plates of $d=1.6$ nm. The
average density in the confined space between hydrophilic plates
is shown in Fig. \ref{nO-tPhi1.0}. As for the hydrophobic plates
case, we define $\rho(P)= \langle n(t) \rangle \times
m_{H2O}/V_{conf}$ and use both $d=1.6$~nm and $P_n(\pm d_{eff})=
0.5$ to compute $V_{conf}$. In contrast to the hydrophobic case,
both methods now lead to very similar $\rho(P)$ values.
 The underestimated (formal volume) and overestimated
(effective volume) values of $\rho(P)$ are very close to each
other and bracket those for bulk water. Interestingly,
Fig.~\ref{nO-tPhi1.0} shows that the slopes of $\rho(P)$ for
confined water (using both volume definitions) are almost the same
as that of bulk water. This suggests that the compressibility of
water confined by hydrophilic plates is the same as that of bulk
water. In fact, we calculate the compressibility as in
Sec.~\ref{vconfSect}, for the case of water confined by
hydrophobic plates. We find that, using either the formal volume
or the effective volume, the compressibility of water confined by
hydrophilic plates is $\kappa_T \approx 0.52$~GPa$^{-1}$, close to
the value $\kappa_T \approx 0.47$~GPa$^{-1}$ of bulk water.

\begin{figure}
\centering{
\includegraphics[width=3.375 in]{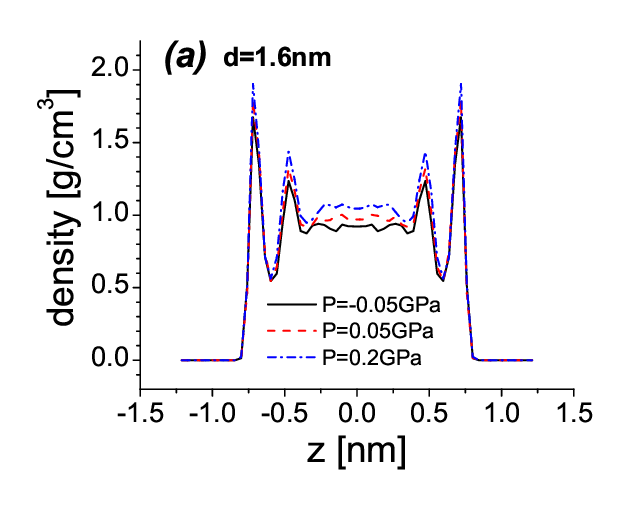}
\includegraphics[width=3.375 in]{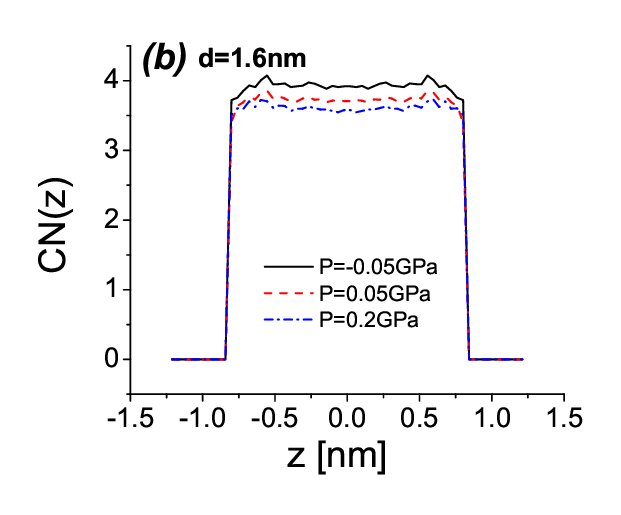}
} \centering{
\includegraphics[width=3.375 in]{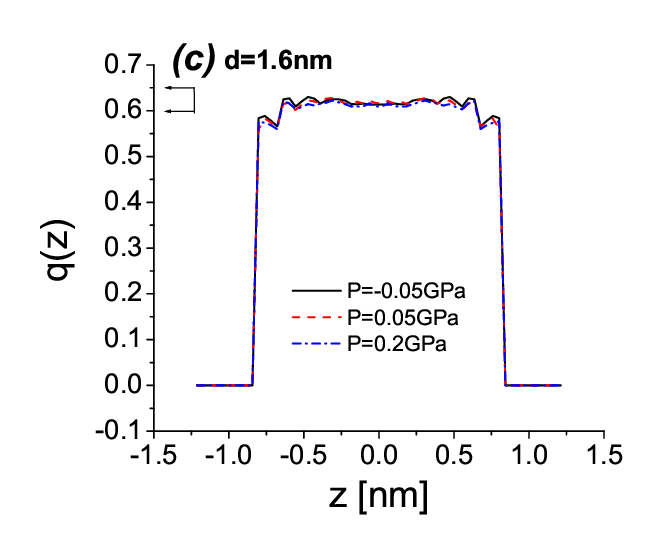}
} \caption{(Color online) Same as Fig.~\ref{X-zPhi0.0}, but for
the case of hydrophilic plates.} \label{X-zPhi1.0}
\end{figure}

Figure \ref{X-zPhi1.0} shows $\rho_{slab}(z)$, $CN(z)$, and $q(z)$
for water confined between hydrophilic plates at different values
of $P$. The three quantities are insensitive to $P$ over the range
of conditions investigated here. Thus, the strong attraction
exerted by the walls on liquid water is not measurably perturbed
by the additional forces associated with compression. The two
maxima of $\rho_{slab}(z)$ in Fig. \ref{X-zPhi1.0}(a) clearly
indicate the presence of two water layers next to the plates. We
note that Fig. \ref{X-zPhi1.0}(a) is very similar to Fig. 3 of
Ref. \cite{lee2} for the TIP4P water model. From Fig.
\ref{X-zPhi1.0}(b) and (c), we find that both $CN(z)$ and $q(z)$
in the presence of hydrophilic plates decrease to zero very
sharply at the plate surfaces. Moreover, a comparison of
Fig.~\ref{X-zPhi0.0}(c) and~\ref{X-zPhi1.0}(c) shows that in the
absence of the hydroxylated groups on the plates, the values of
$q(z)$ start to decrease at larger distances from the walls than
in the case of hydrophilic plates. The resulting profiles in
Fig.~\ref{X-zPhi1.0}(b) and (c) are almost flat for the region
between the plates meaning that the values of $CN(z)$ and $q(z)$
next to the plates barely change with respect to their bulk water
values at $z=0$. These results suggest that water molecules far
from the plates have similar local environment (in terms of
tetrahedral order and number of nearest-neighbors) as those next
to the plates (when including the plate surface oxygen atoms). For
the molecules next to the plates, the $O-H$ atoms on the plate
surface act as `virtual' water molecules providing extra HBs to
the real water molecules. Again, the slight change in $CN(z)$ at
$z \approx 0$ is probably due to the fact that we use a fixed
cutoff when calculating the coordination number.

\begin{figure}
\includegraphics[width=3.375 in]{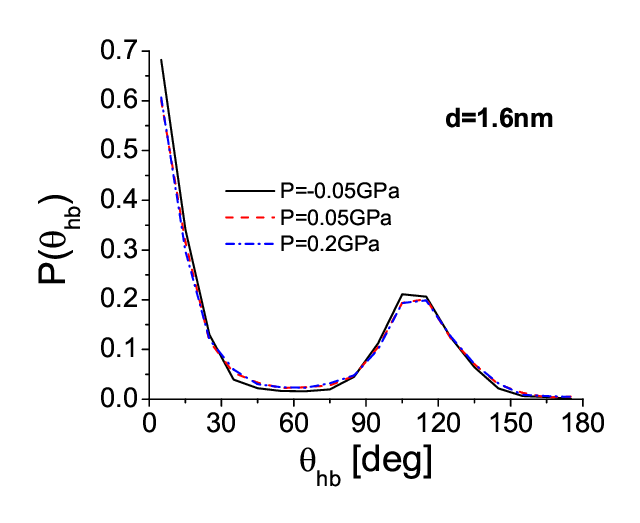}
\caption{(Color online) Normalized distribution $P(\theta_{hb})$
of angles $\theta_{hb}$ between the hydrogen bond vectors of water
molecules and the normal direction to the plates, for molecules at
a distance $\leq 0.1$~nm from the plates
%$P(\theta_{hb})$ shows
%maxima at $\theta_{hb}= 0^o$ and $110^o$ and is the complementary
%distribution found in the case of hydrophobic plates
 ({\it cf} Fig.~\ref{TitaHbPhi0.0}).  Hydrophilic plates. } \label{TitaHbPhi1.0}
\end{figure}

Figure \ref{TitaHbPhi1.0} shows $P(\theta_{hb})$ corresponding to
a slab next to the plates of width $0.1$~nm. $P(\theta_{hb})$ is
in agreement with the corresponding distribution computed in ref.
\cite{lee2} for the TIP4P water model and is complementary to the
distribution found in the case of hydrophobic plates (see Fig.
\ref{TitaHbPhi0.0}). The presence of polarity on the plate allows
water molecules to form HBs with atoms on the plate. The resulting
$P(\theta_{hb})$ increases abruptly for $\theta_{hb}<20^o$ and has
a sharp peak at $\theta_{hb} \approx 110^o$. As already found in
\cite{lee2} for the TIP4P water model, these maxima can be
explained by the presence of multiple HBs that a water molecule
can have with $OH$ groups associated with the plate (see Fig. 14
in \cite{lee2}). Ultrafast electron crystallography measurements
of water confined between hydrophilic walls \cite{ocho} also show
that water molecules interact at two sites of the substrate.

\begin{figure}
\centering{
    \includegraphics[width=2.5 in]{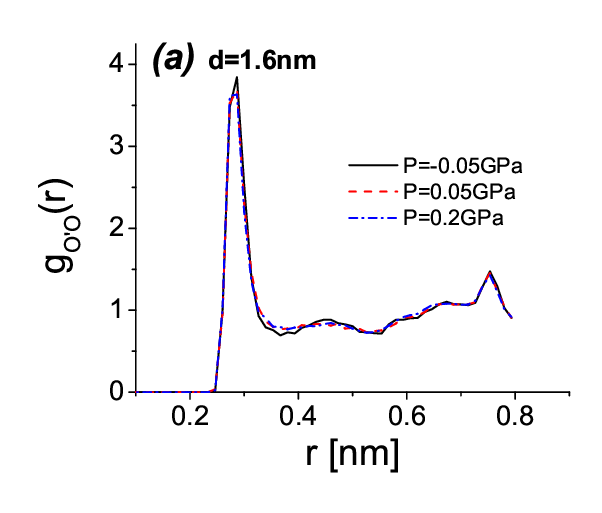}
    \includegraphics[width=2.5 in]{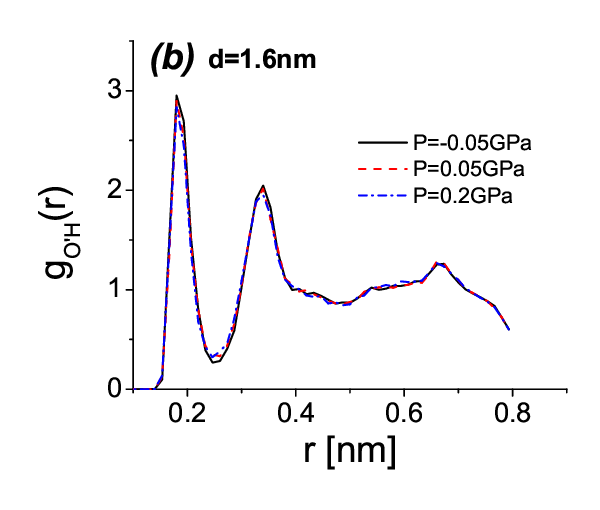}
    }
\centering{
    \includegraphics[width=2.5 in]{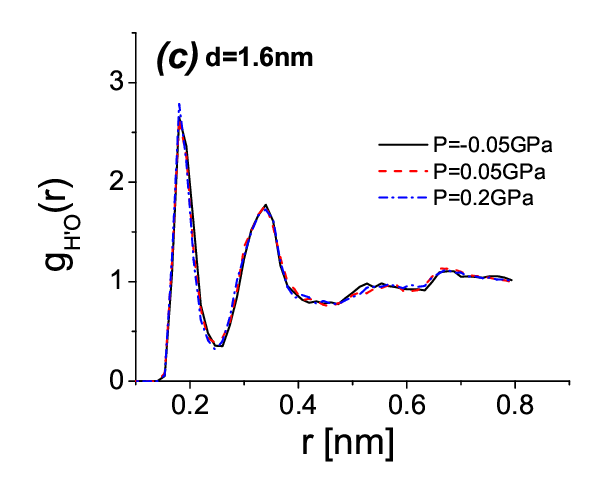}
    \includegraphics[width=2.5 in]{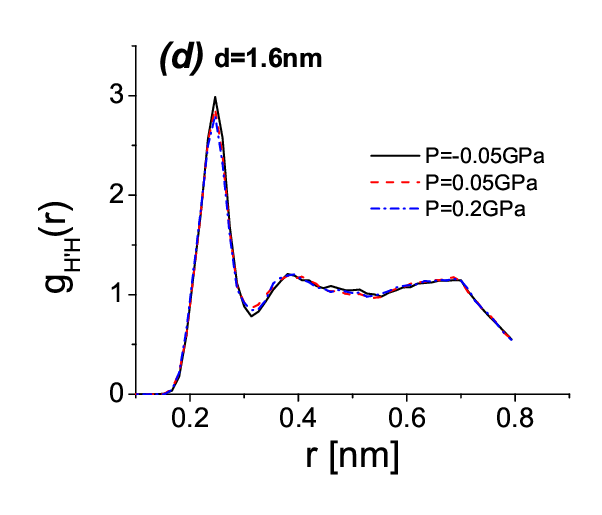}
    }
\caption{(Color online) Radial distribution functions for the
plate oxygen atoms ($O'$), and (a) the water oxygen and (b) water
hydrogen atoms. (c)(d) Same as (a), (b) for the plate hydrogen
atoms ($H'$). Hydrophilic plates.} \label{rdfPhi1.0}
\end{figure}

The RDFs between the plate oxygen ($O'$) or hydrogen ($H'$) atoms
and the water molecule oxygen atoms are shown in Fig.
\ref{rdfPhi1.0}. These distributions do not change with $P$,
confirming the insensitivity of confined water structure between
hydrophilic plates to changes in external pressure.
Figure~\ref{rdfPhi1.0} is in agreement with Figs. 11 and 12 of
Ref. \cite{lee2}, which were obtained for the TIP4P water model.

\section{Summary}
\label{summary}

We have presented results from MD simulations of water confined
between nanoscale (hydrophobic or hydrophilic) plates at $T=300$~K
and for a range of values of pressure $P$ ($-0.15$~GPa~$\leq P
\leq 0.2$~GPa) and plate-plate separation $d$ ($0.4$~nm~$\leq d
\leq 1.6$~nm).

 In the case of hydrophobic plates, a phase diagram in
the $P-d$ plane summarizing the MD results is presented, and three
phases (liquid, vapor, and bilayer ice) are identified. At low $P$
we find capillary evaporation. The distance at which this drying
occurs decreases with increasing $P$. Furthermore, the transition
line in the $P-d$ phase diagram separating the vapor and liquid
phases is followed to small values of $d$, below which water
molecules become individually trapped and immobilized by the
surface atoms.
 It is possible that at higher
$T$, this line ends in a critical point at high $d$ (corresponding
to the liquid-vapor critical point of bulk water).

The bilayer ice is composed of two layers of hexagons in registry
 along the surface normal.
The resulting ice is similar to that found in
Ref.~\cite{KogaIcePrl}. This ice is found at $P\geq 0.1$~GPa and
$d\approx 0.6$~nm, but only in a narrow range of $d$. The region
corresponding to ice in the $P-d$ phase diagram might be connected
with the vapor-liquid transition line for larger plates,
corresponding to a triple point. However, the plates we simulate
are small, and we cannot observe simultaneously the liquid, vapor
and crystal phases. Further investigation is required on this
point. We also note that at $T=300$~K and in the presence of
nanoscale plates, we do not find indications of a monolayer ice,
like that observed in MD simulations using the TIP4P and TIP5P
potentials and infinite walls \cite{zangi1}.
%This can be due to the simple
%potential (SPC/E) we use or to the high $T$ we simulate
%\cite{tanaka2005}.

When simulating water confined by hydrophilic (hydroxylated
silica) plates, we find that the confined water remains in the
liquid phase at all $P$ and $d$ studied. Moreover, even at $P$
where we observe cavitation in the bulk water, the confined water
is in the liquid phase. In other words, the hydrophilic plates
induce wetting in the confined space (a phenomenon reminiscent of
capillary condensation). This result suggests that one way to
stabilize liquid water under tension may be by hydrophilic
confinement. Experiments of water confined between a hydrophobic
and a hydrophilic surface (Janus interface) \cite{seis}, suggest a
similar physical picture, where the hydrophobic surface encourages
water to dewet, while the hydrophilic surface constrains water to
be present.

We also study the effect of $P$ on the hydration of both
hydrophobic and hydrophilic plates. We focus on the water molecule
distribution, average coordination number and local tetrahedral
order parameter along the direction normal to the plates. In the
case of hydrophobic plates, all of these quantities indicate that
as $P$ increases, water molecules are pushed toward the plates
while the liquid becomes more structured. However, the water
molecule orientation next to the plates shows no dependence on
$P$. As previously reported in ref. \cite{lee1}, molecules have on
average one HB pointing to the plate. Strikingly, for the case of
hydrophilic plates, we find no change in the liquid structure with
$P$ in the range $-0.1$~GPa to $0.2$~GPa.

The strong differences observed in the behavior of water confined
 between hydrophilic and hydrophobic plates leads to the
 interesting question of what the thermodynamic and structural properties
 of water are when it is confined between {\it heterogeneous} plates (e.g.,
 hydrophobic plates that have been partially hydroxylated).
 Results addressing this question will be reported in a subsequent report.

\section*{Acknowledgments}

We thank F. W. Starr for providing the basic SPC/E code for water
using the reaction field method, M.S. Shell for fruitful
discussions related to the Ewald sum method, and
 F. Sciortino for enlightening discussions related to the simulation
 details. We also thank A. Luzar for very useful comments on the manuscript.
 PJR gratefully acknowledges support by the R. A.
 Welch Foundation (F-0019). PGD and PJR gratefully acknowledge the support of the
 National Science Foundation (Collaborative Research in Chemistry
 Grant Nos. CHE$0404699$ and CHE$0404695$).

\newpage

\vspace{1cm}
\begin{table}[!h]
\caption{Potential parameters for plate-water interactions (taken
from Ref.~\cite{lee2}).\label{tabla} }
\begin{tabular}{c c c c}
 \hline \hline
 Atom Type & $\epsilon$ [kJ/mol]$^{(a)}$
& $\sigma$ [nm]$^{(a)}$ & Charge [e]$^{(b)}$ \\
 \hline
 $O$ &  $0.6487$ & $0.3154$ & $-0.71$ \\
  $Si$ &  $0.5336$ & $0.3795$ & $0.31$ \\
  $H$  &  $-$   &  $-$  & $0.40$ \\
\hline \hline
\end{tabular}\\
\raggedright {$^{(a)}$ Lennard-Jones parameter for plate-water
$O$ atom interactions (hydrophobic and hydrophilic plates).\\
$^{(b)}$ Charges on $Si-O-H$ groups (hydrophilic plates).}
\end{table}
\vspace{1cm}

\vspace{1cm}
\begin{table}[!h]
\caption{Simulation times for the different kinds of plates and
plate separation $d$. \label{tabla2}}
\begin{tabular}{c c c}
 \hline \hline
 Plate &$d$ [nm] &
 Simulation time [ps] \\
 \hline
   hydrophobic &  $>0.8$ & $200$ \\
   hydrophobic &  $\leq 0.8^{(a)}$ &  $400^{(a)}$ \\
   hydrophilic & $\leq 1.6$  & $200$ \\
\hline \hline
\end{tabular}\\
\raggedright{$^{(a)}$ For hydrophobic plates with $d=0.6$~nm and
$P=0.05$~GPa, the simulation time is $1$~ns.}
\end{table}
\vspace{1cm}

\newpage

\newpage

\newpage

\newpage

\newpage

\newpage

\newpage

\newpage

\newpage

\newpage

\newpage

\newpage

\newpage

\newpage

\begin{thebibliography}{23}
\bibitem{footnoteLayer}
The presence of the small third peak in Fig.~\ref{msdIce}c is due
to crystal defects occurring during the ice formation
(approximately $t<300$~ps), which causes some diffusing molecules
between the two ice layers during the transient crystallization to
be included in the statistics. Furthermore, there are "border
effects" due to the finite size of the plates. Thus, molecules at
the boundaries of the plates (which do not belong to the crystal)
might have been included in the statistics.
\end{thebibliography}
\end{document}